\pdfoutput=1

\documentclass[twocolumn,showpacs,amssymb,aps,nofootinbib,floatfix,superscriptaddress]{revtex4-1}

\usepackage{epsfig}
\usepackage{float}
\usepackage{amssymb}
\usepackage{hyperref}

\newcommand{\be}{\begin{equation}}
\newcommand{\ee}{\end{equation}}

\newcommand{\bea}{\begin{eqnarray}}
\newcommand{\eea}{\end{eqnarray}}
\newcommand{\ba}{\begin{array}}
	\newcommand{\ea}{\end{array}}

\begin{document}

\title{Correlation coefficient  between harmonic flow and transverse  momentum in heavy-ion collisions}

\author{Piotr Bo\.zek}
\email{Piotr.Bozek@fis.agh.edu.pl}
\affiliation{AGH University of Science and Technology, Faculty of Physics and
Applied Computer Science, aleja Mickiewicza 30, 30-059 Cracow, Poland}

\author{Hadi Mehrabpour}
\email{hadi.mehrabpour@ipm.ir}
\affiliation{School of Particles and Accelerators, Institute for Research in Fundamental Sciences (IPM), P.O. Box 19395-5531, Tehran, Iran }
\affiliation{Department  of  Physics,  Sharif  University  of  Technology,  P.O.  Box  11155-9161,  Tehran,  Iran} 

\begin{abstract}
The correlation between the harmonic flow and the 
transverse momentum in relativistic heavy ion collisions is  calculated in the hydrodynamic model. The partial correlation coefficient, corrected for fluctuations of multiplicity, is compared to experimental data. Estimators of the final transverse momentum and harmonic flow are used
to predict the value of the correlation coefficient from the moments of 
the initial distribution. A good description of the hydrodynamic simulation 
results 
is obtained if the estimator
 for the final transverse momentum, besides the transverse
 size and the entropy, includes also the eccentricities.
\end{abstract}


\maketitle

\section{Introduction}

The dynamics of relativistic heavy ion  collisions is studied experimentally by measuring  characteristics of particles emitted in collision events. Some of the most common observables used in heavy ion collisions are the harmonic flow coefficients, measuring the azimuthal asymmetry of the emitted hadrons, and  
transverse momentum spectra. In the hydrodynamic scenario these two quantities
are a measure of  the
collective expansion of the dense matter created in the interaction region \cite{Ollitrault:2010tn,Heinz:2013th,Gale:2013da}.

In order to find an additional characteristic of the rapid expansion, a
 correlation measurement between the harmonic flow and 
transverse momentum has been   proposed \cite{Bozek:2016yoj}.
In this paper we present results for the harmonic flow-transverse momentum 
correlation coefficient in Pb+Pb and p+Pb collisions 
at $\sqrt{s_{NN}}=5.02$~TeV.
Experimental results for these collisions have been published by the 
ATLAS Collaboration \cite{Aad:2019fgl}. 
The calculated correlation coefficients 
are corrected for effects of multiplicity fluctuations within each centrality 
bin, using the method  of partial correlation coefficients
 \cite{Olszewski:2017vyg}.

 The values of the final global collective variables, such as the 
 harmonic flow coefficients and the average transverse momentum, can be
 reasonably well estimated from the initial entropy, 
transverse size and eccentricities.
We study how well such estimators of the final observables
 predict the correlation coefficient
 between the  final harmonic flow and the transverse momentum. 
Linear hydrodynamic response is superimposed on moments of the initial
 density to calculate the covariances between the final observables.

\section{Model}

The collision dynamics  is described by the viscous hydrodynamic model \cite{Bozek:2009dw,Schenke:2010rr}.
The initial entropy density in the transverse plane is generated from the nucleon Glauber model. Each participant nucleon contributes to the initial entropy of the fireball.  The system is evolved by the hydrodynamic equations  with shear viscosity $\eta/s=0.08$ and a temperature dependent bulk viscosity \cite{Bozek:2011ua}. At the freeze-out temperature of $150$~MeV hadrons are emitted statistically \cite{Chojnacki:2011hb}.  We perform simulation for Pb+Pb and p+Pb collisions at  $\sqrt{s_{NN}}=5.02$ TeV. Details of the calculation can be found in Refs.  \cite{Bozek:2011ua,Bozek:2013uha}.

The azimuthal spatial anisotropies of the initial entropy density profile $s(r,\phi)$ in the transverse plane 
are characterized by the eccentricities
\begin{equation}
\epsilon_n e^{i n \Psi_n} = - \frac{\int rdr d\phi \  r^{n} s(r,\phi )e^{i n \phi  }}{\int rdr d\phi  \ r^{n} s(r,\phi )} \ .
\end{equation} 
The hydrodynamic evolution of an azimuthally asymmetric distribution leads to an an azimuthal asymmetry in particle spectra. 
For $N$  particles emitted  in the acceptance region the harmonic flow coefficients are calculated
\begin{equation}
v_n\{2\}^2=  \frac{1}{N(N-1)} \sum_{j\neq k=1}^{N}   e^{\i n (\phi_j-\phi_k)}
\end{equation} 
in each event. The average transverse momentum in each event is defined as
\begin{equation}
[p_T]=\frac{1}{N} \sum_{i=1}^{N} p_{i } \ .
\end{equation}
The flow coefficients and the average transverse momentum are  calculated for charged particles in most of the cases, but
we present also some 
results for identified particles, protons, kaons, and pions. 
To improve the statistics we use combined events generated from the same hydrodynamic evolution. This procedure allows to reduce the statistical error in correlations an the corrections for selfcorrrelations can be neglected \cite{Bozek:2016yoj}.

\section{Partial Correlation}

Covariances and variances  of observables  in heavy-ion collisions are measured in experiments and 
predicted in model calculations \cite{Borghini:2001vi,Bilandzic:2010jr,Jia:2014jca,Qiu:2012uy,Bhalerao:2013ina}. In most cases these are quantities based on covariances (or cumulants) of flow coefficients.
In this paper we consider the correlation between the harmonic flow coefficients and the average transverse momentum
\begin{equation}
\rho(v_n\{2\}^2,[p_T])=\frac{Cov(v_n\{2\}^2,[p_T])}{\sqrt{Var(v_n\{2\}^2)Var([p_T])}}  \ .
\label{eq:rhovnp}
\end{equation}
The covariances and variances in the above formula should be calculated excluding self-correlations, 
i.e. the sums over many particles should be done excluding same particle indices \cite{Bozek:2016yoj}, i.e.
with 
\begin{eqnarray}
Cov(v_n\{2\}^2,[p_T]) & = & \langle\frac{1}{N(N-1)(N-2)} \nonumber  \\ 
& & \sum_{i\neq j\neq k} e^{ i n(\phi_i-\phi_j)}\left( p_{k}- \langle [p_T]\rangle \right)\rangle  
\label{eq:covvnpt}
\end{eqnarray}
and using dynamical variances
\begin{eqnarray}
 Var(v_{n}^2)_{dyn} &  = & \nonumber \\ & & \langle \frac{1}{N(N-1) (N-2)(N-3)}  \nonumber \\ 
& &  \sum_{i \neq j  \neq k \neq l  }  e^{in\phi_i+in \phi_j} e^{-in\phi_k-i n \phi_l}  \rangle \nonumber \\
&  & - \langle \frac{1}{N (N-1)} \sum_{i \neq k  }  e^{in\phi_i} e^{-in\phi_k} \rangle^2  \ 
\label{eq:varvn}
 \end{eqnarray}
%\begin{equation}
%Var(v_n\{2\}^2)=v_n\{2\}^4-v_n\{4\}^4 \ 
%\end{equation}
and
\begin{equation}
Var([p_T])=\langle \frac{1}{N(N-1)}\sum_{i\neq j} (p_i -\langle [p_T]\rangle )(p_j -\langle [p_T]\rangle )\rangle \ ,
\label{eq:varpt}
\end{equation} 
$\langle \dots \rangle$ represents the average over events.
If the correlations between emitted particles come from the collective flow only, the estimators in Eqs. \ref{eq:covvnpt}, \ref{eq:varvn}, and \ref{eq:varpt} represent the covariance and the variances of the respective collective variables, with  statistical fluctuations removed.
Predictions for the harmonic flow-transverse momentum correlation have been presented previously for Pb+Pb collisions
 at $\sqrt{s_{NN}}=2760$~GeV \cite{Bozek:2016yoj}. In the following we present hydrodynamic model results for Pb+Pb collisions 
for a higher energy, corresponding to the data published by the ATLAS Collaboration \cite{Aad:2019fgl}.

A direct comparison of the calculation to data is not possible if the centrality bins in the experiment and in the model calculation are different. The ATLAS data are obtained in very narrow multiplicity bins, whereas model calculations are done in relatively broad centrality bins, 5\% or 10\%. In a given centrality bin the multiplicity fluctuates  and such fluctuations may influence the measurement of the flow-transverse momentum correlation.

\begin{figure}[ht]
%  	\begin{center}
  		\begin{tabular}{c}
  			\includegraphics[scale=0.3]{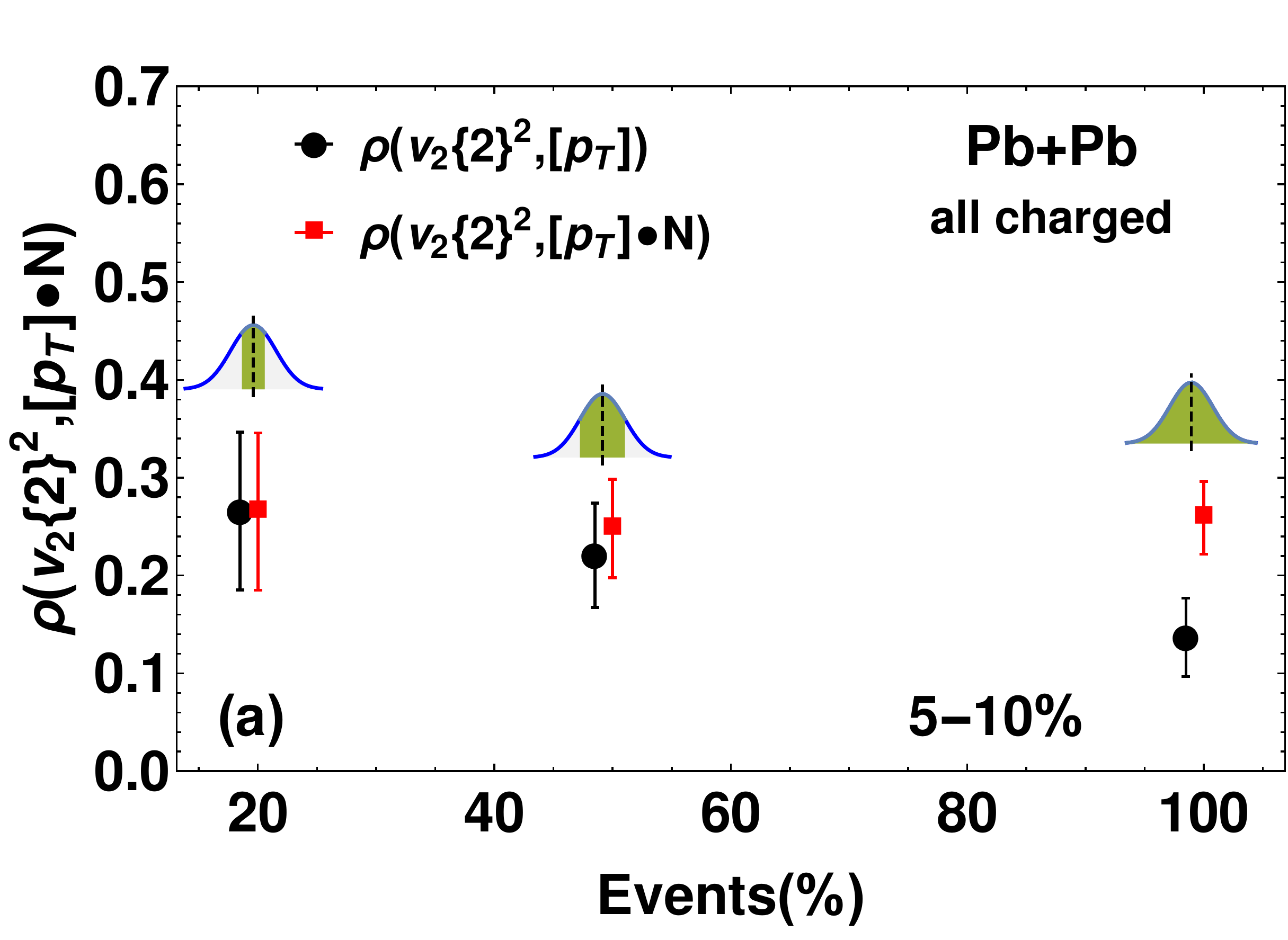} \\	       				\includegraphics[scale=0.3]{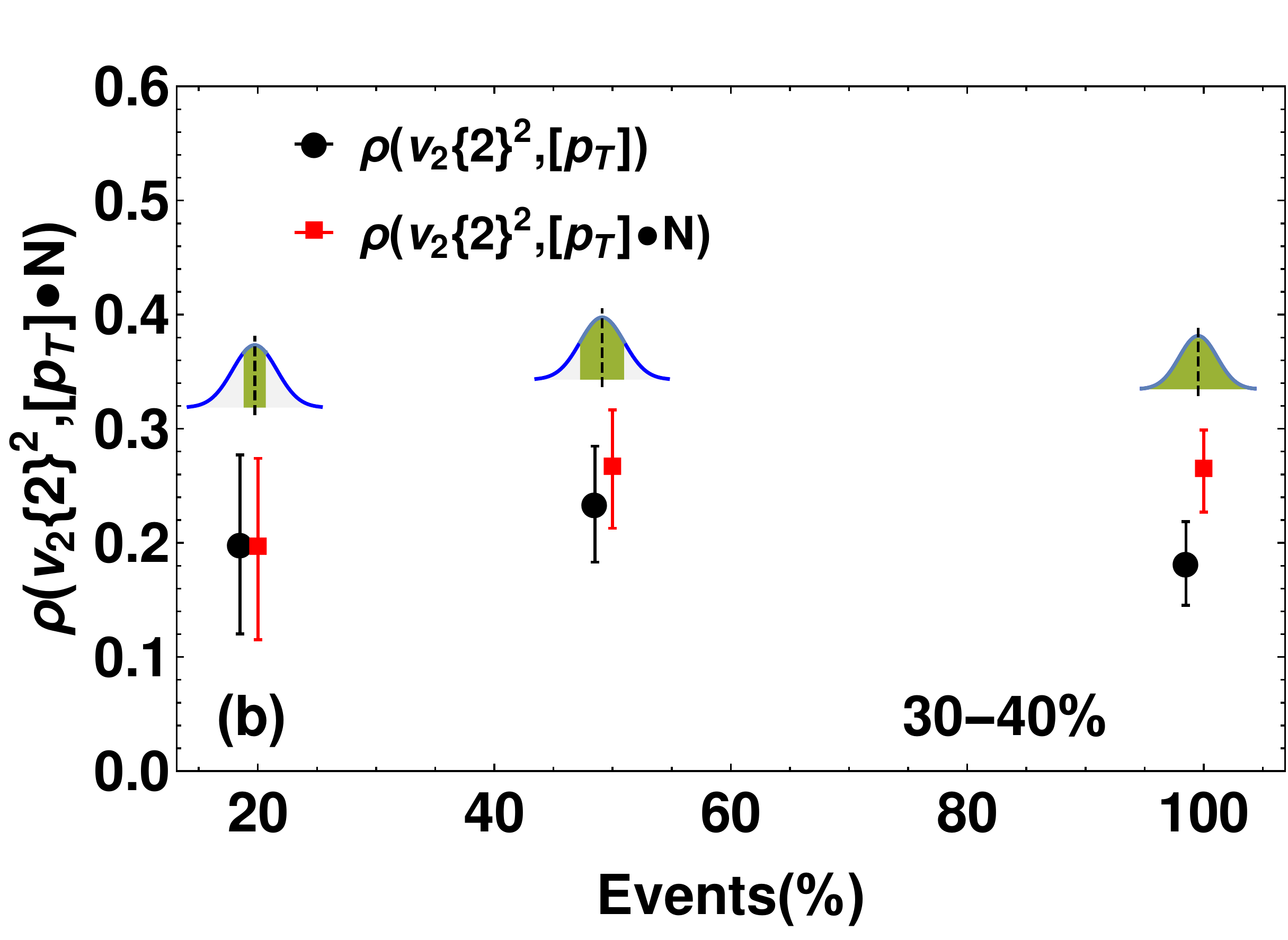} 
  		\end{tabular}		
  		\caption{ The harmonic flow-transverse momentum correlation coefficient (black dots) compared to the 
partial correlation coefficient (red squares) for three different width of the multiplicity bins, all
events (100\%), 50\% of   events, and 20\% of   events  cut out  from the multiplicity distribution.
Panel (a) is for the 5-10\% and panel (b) is for the 30-40\% centrality bin.}
  	\label{fig:partialconverge}
%  \end{center}
\end{figure}

This effect in the context of heavy-ion collisions is discussed in Ref. \cite{Olszewski:2017vyg}. The problem  is how to extract the correlation between two physical observables without interference from a third, control variable.
In our case it is the question how to extract the correlation between the flow harmonic $v_n\{2\}^2$ and the
 average transverse momentum $[p_T]$, without interference due to changes in the control variable, the event multiplicity $N$.
The most direct way is to fix the control variable and to 
calculate all statistical averages in an ensemble of events with fixed multiplicity. This would give the conditional 
correlation coefficient at fixed multiplicity
\begin{equation}
\rho(v_n\{2\}^2,[p_T]|N)=\frac{Cov(v_n\{2\}^2,[p_T]|N)}{\sqrt{Var(v_n\{2\}^2|N)Var([p_T]|N)}}  \ .
\end{equation}
The experimental data is calculated in narrow bins of multiplicity approximating the above procedure  \cite{Aad:2019fgl}.
An alternative way to estimate the correlation coefficient at fixed multiplicity is to use the partial correlation coefficient with correction for effects due to fluctuations in the control variable \cite{Olszewski:2017vyg}.
Using the partial covariance
\begin{eqnarray}
& & Cov(v_n\{2\}^2,[p_T]\bullet N)=  Cov(v_n\{2\}^2,[p_T])\nonumber \\
& & -\frac{Cov(v_n\{2\}^2, N)Cov(N,[p_T] )}{Var(N)}
\end{eqnarray}
and the partial variances
\begin{equation}
Var(v_n\{2\}^2\bullet N)=Var(v_n\{2\}^2)-\frac{Cov(v_n\{2\}^2, N)^2}{Var(N)} \ ,
\end{equation}
\begin{equation}
Var([p_T]\bullet N)=Var([p_T])-\frac{Cov([p_T], N)^2}{Var(N)} \ ,
\end{equation}
one gets for partial correlation coefficient
\begin{eqnarray}
& & \rho(v_n\{2\}^2,[p_T]\bullet N) = \nonumber\\ & &\frac{Cov(v_n\{2\}^2,[p_T]\bullet N)}{\sqrt{Var(v_n\{2\}^2\bullet N)Var([p_T]\bullet N)}} =\nonumber \\
& & \frac{ \rho(v_n\{2\}^2,[p_T])-\rho(v_n\{2\}^2, N)\rho(N,[p_T] )}{\sqrt{1-\rho(v_n\{2\}^2, N)^2}\sqrt{1-\rho(N,[p_T] )^2}} \ .
\end{eqnarray}

The application of the partial correlation analysis is illustrated in Fig. \ref{fig:partialconverge}.
The standard correlation coefficient $\rho(v_n\{2\}^2,[p_T])$ is calculated for three different ensembles of events 
width full and reduced  width of the multiplicity distribution (black dots)\footnote{The shape of multiplicity distribution is approximately a Gaussian distribution in all centrality classes. The centrality bins in the simulation are defined by the number of participants, not the final multiplicity.}. In the limit of  zero width, one would recover
 the correlation coefficient at fixed multiplicity. In practice, we stop at an ensemble with $20\%$ of events from the  center of the multiplicity distribution, due to limited statistics.
 One notices that the results depend on the width of the multiplicity bin. The limit of fixed multiplicity can be estimated using the partial correlation coefficient  $\rho(v_n\{2\}^2,[p_T]\bullet N)$ (red squares in Fig. \ref{fig:partialconverge}). With general assumptions, it is expected that the partial correlation coefficient does
 not depend on  
the width of the multiplicity bin \cite{Olszewski:2017vyg}. In our numerical results it true within the statistical uncertainities.
 The two correlation coefficients merge in the most narrow multiplicity bin,
 although with increasing a large error.

\section{Partial correlation analysis of flow and transverse momentum}

%\subsection{Charged particles}

\begin{figure}[ht]
%  	\begin{center}
  		\begin{tabular}{c}
  			\includegraphics[scale=0.3]{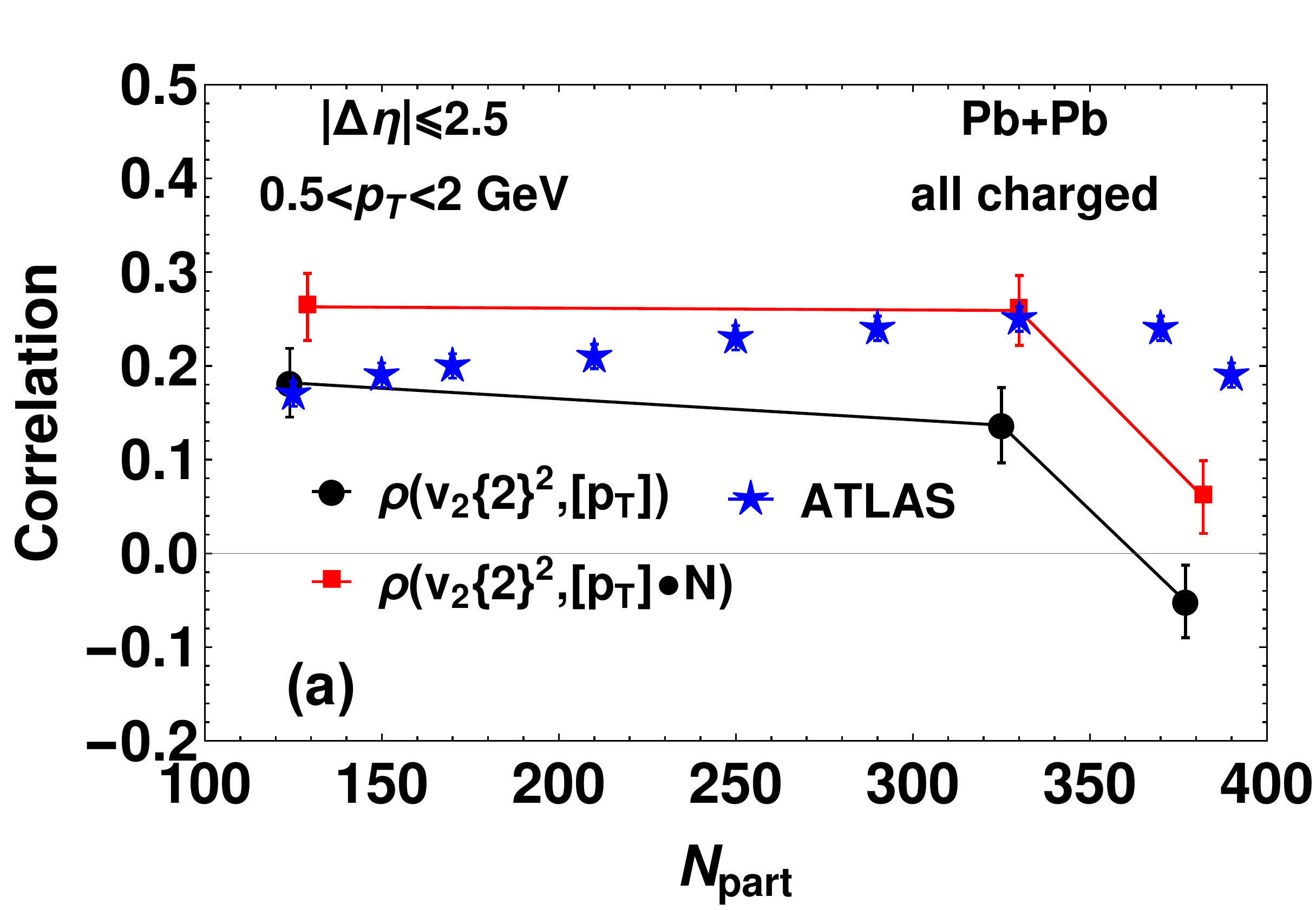} \\	       				\includegraphics[scale=0.3]{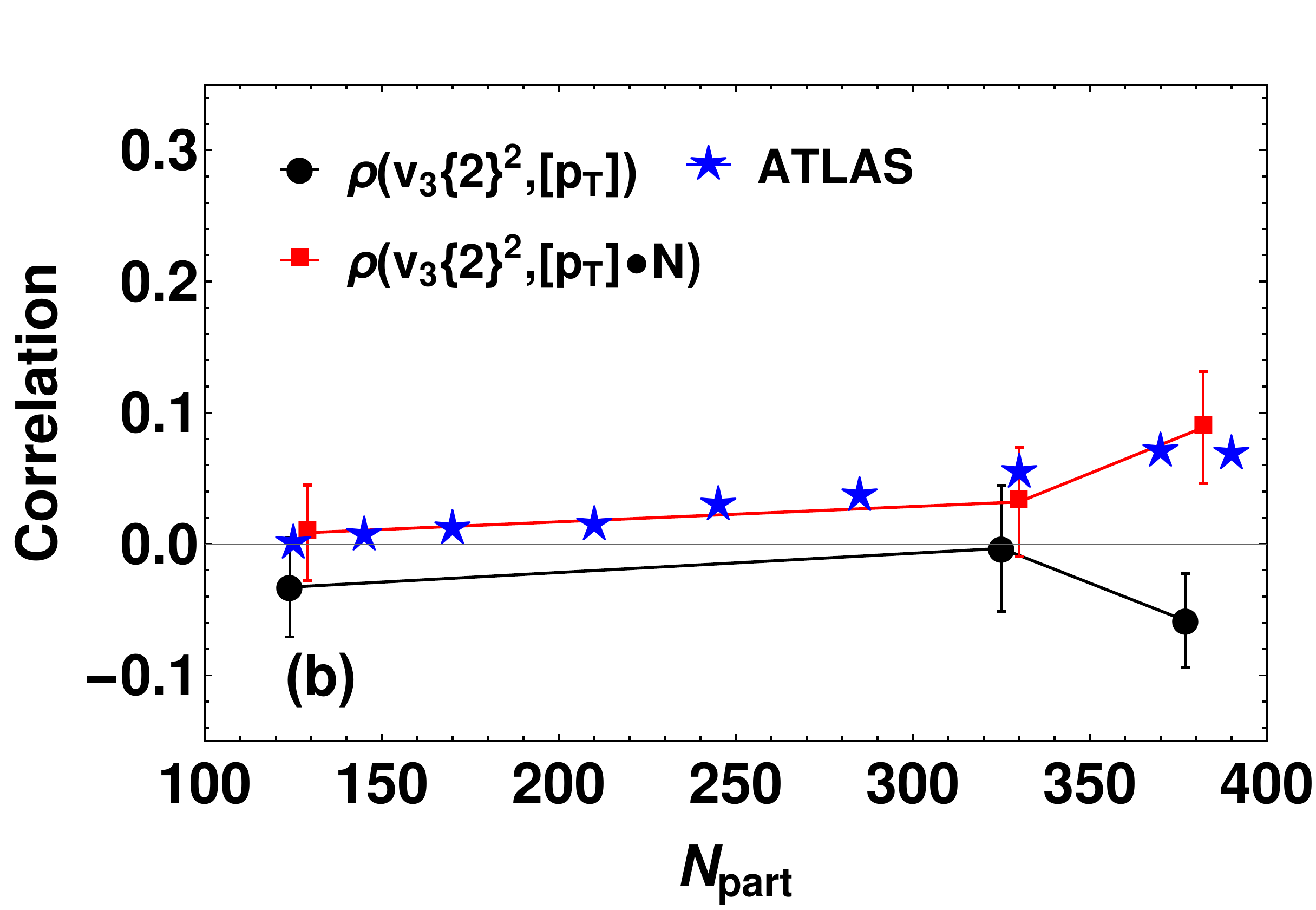} 
  		\end{tabular}		
  		\caption{ The harmonic flow-transverse momentum correlation coefficient (black dots) and the 
partial correlation coefficient (red squares) compared to ATLAS Collaboration data (blue stars) \cite{Aad:2019fgl}  for Pb+Pb collisions as a function the number of participant nucleons.
Panels (a) and  (b) are for elliptic and triangular flow respectively.
  	\label{fig:vnpt}}
%  \end{center}
\end{figure}

We calculate the correlations coefficient and the partial correlation coefficient for charged hadrons emitted in Pb+Pb collisions at $\sqrt{s_{NN}}=5.02$TeV. The 
results obtained in the hydrodynamic model are shown in Fig. \ref{fig:vnpt}.
We note that the corrections due multiplicity fluctuations are significant,
the partial correlation coefficient $ \rho(v_n\{2\}^2,[p_T]\bullet N)$ 
is larger that the standard correlation coefficient  $\rho(v_n\{2\}^2,[p_T])$.
The experimental data are taken in narrow bins of centrality and approximate
the  correlation coefficient  at fixed multiplicity $\rho(v_n\{2\}^2,[p_T]| N)$.
The calculated partial correlation coefficient fairly well reproduces the measured data
both for the elliptic and triangular flow.

\begin{figure}[t!]
%  	\begin{center}
  		\begin{tabular}{c}
  			\includegraphics[scale=0.3]{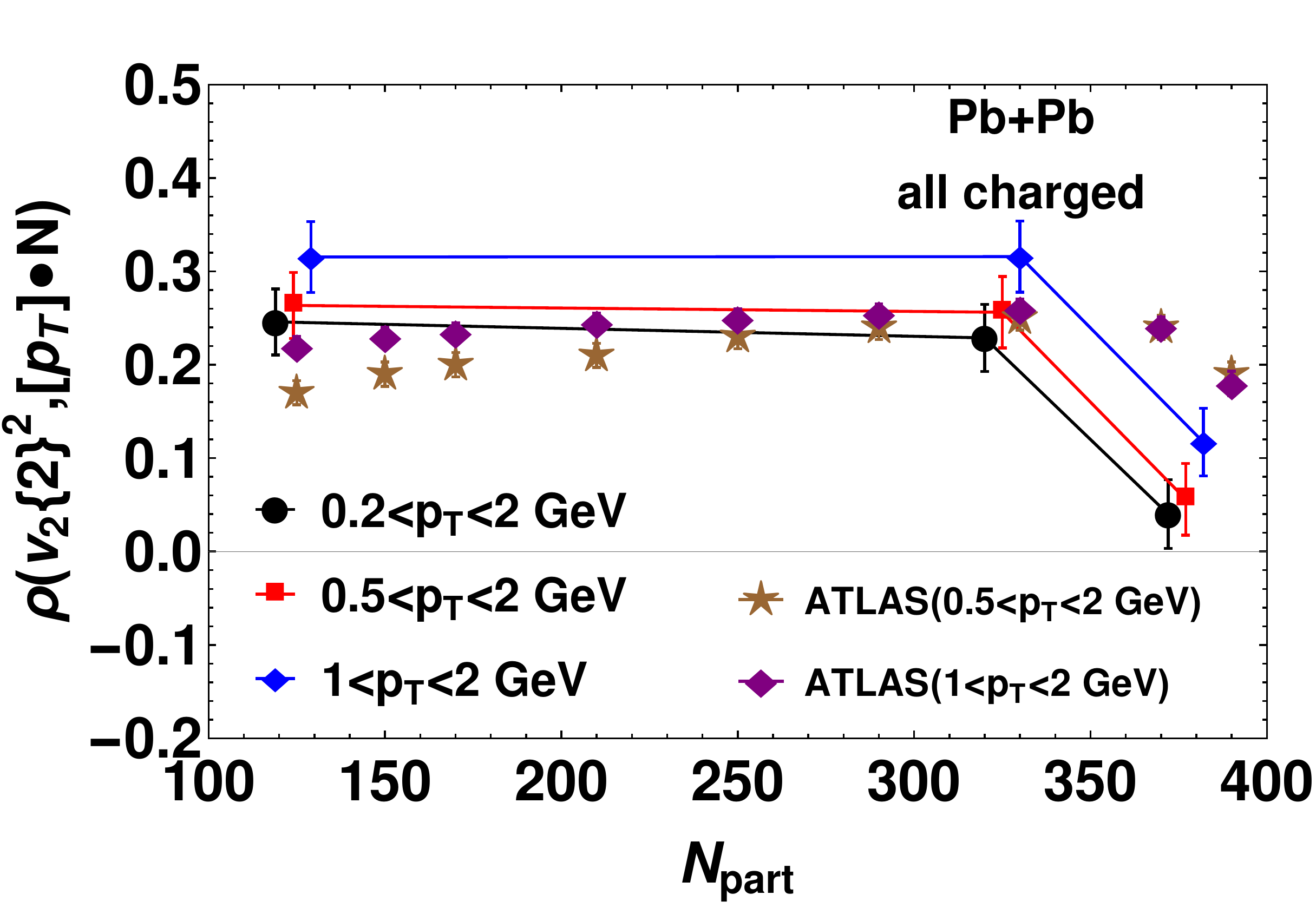} 	       
  		\end{tabular}		
  		\caption{The elliptic flow-transverse momentum correlation 
coefficient for charged particles emitted in Pb+Pb collisions for different $p_T$ cuts, $0.2<p_T<2$~GeV (black dots), $0.5<p_T<2$~GeV (red squares), and $1<p_T<2$~GeV (blue diamonds),  }
  	\label{fig:ptcut}
%  \end{center}
\end{figure}

\begin{figure}[ht]
%  	\begin{center}
  		\begin{tabular}{c}
  			\includegraphics[scale=0.3]{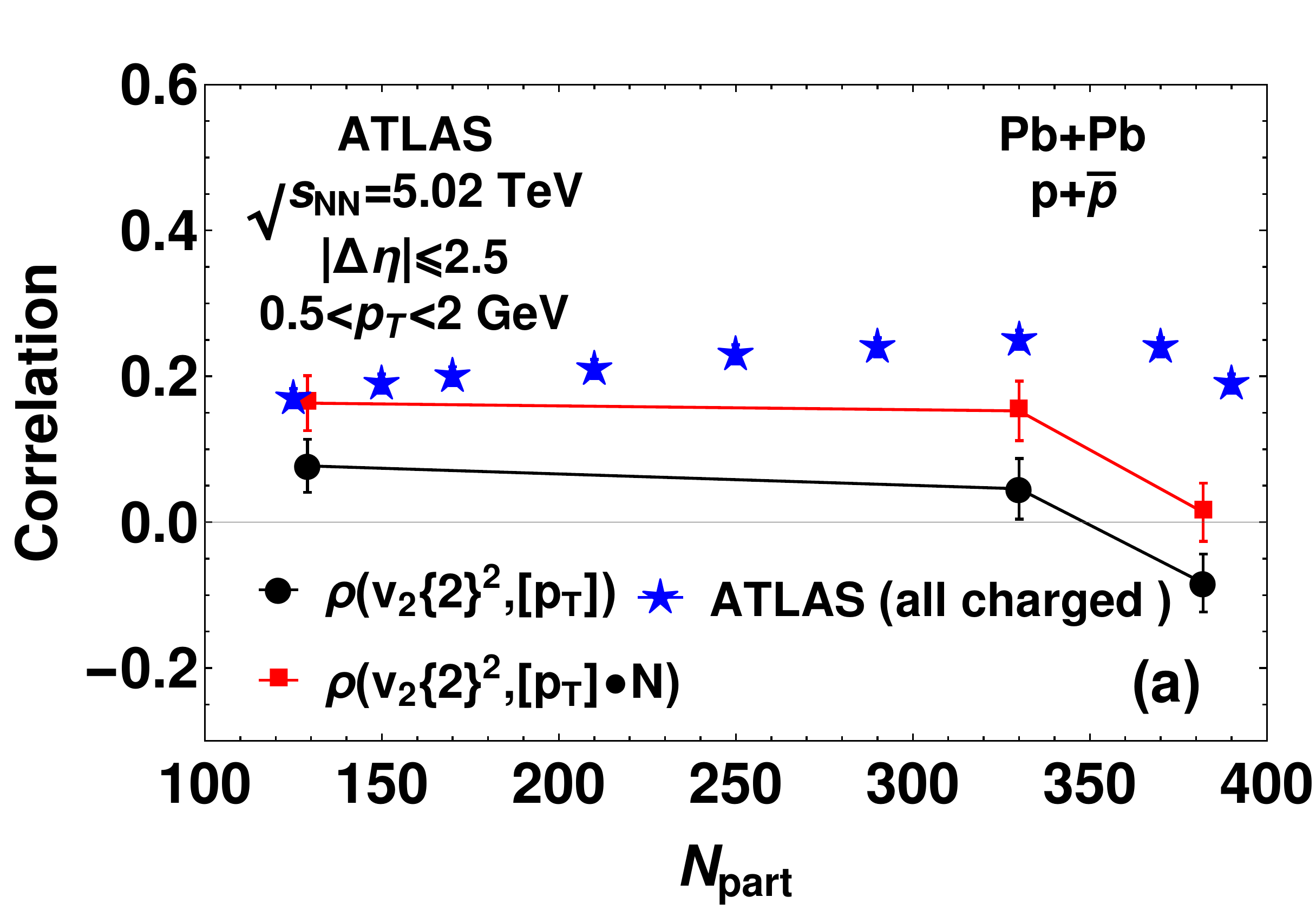} \\	       				\includegraphics[scale=0.3]{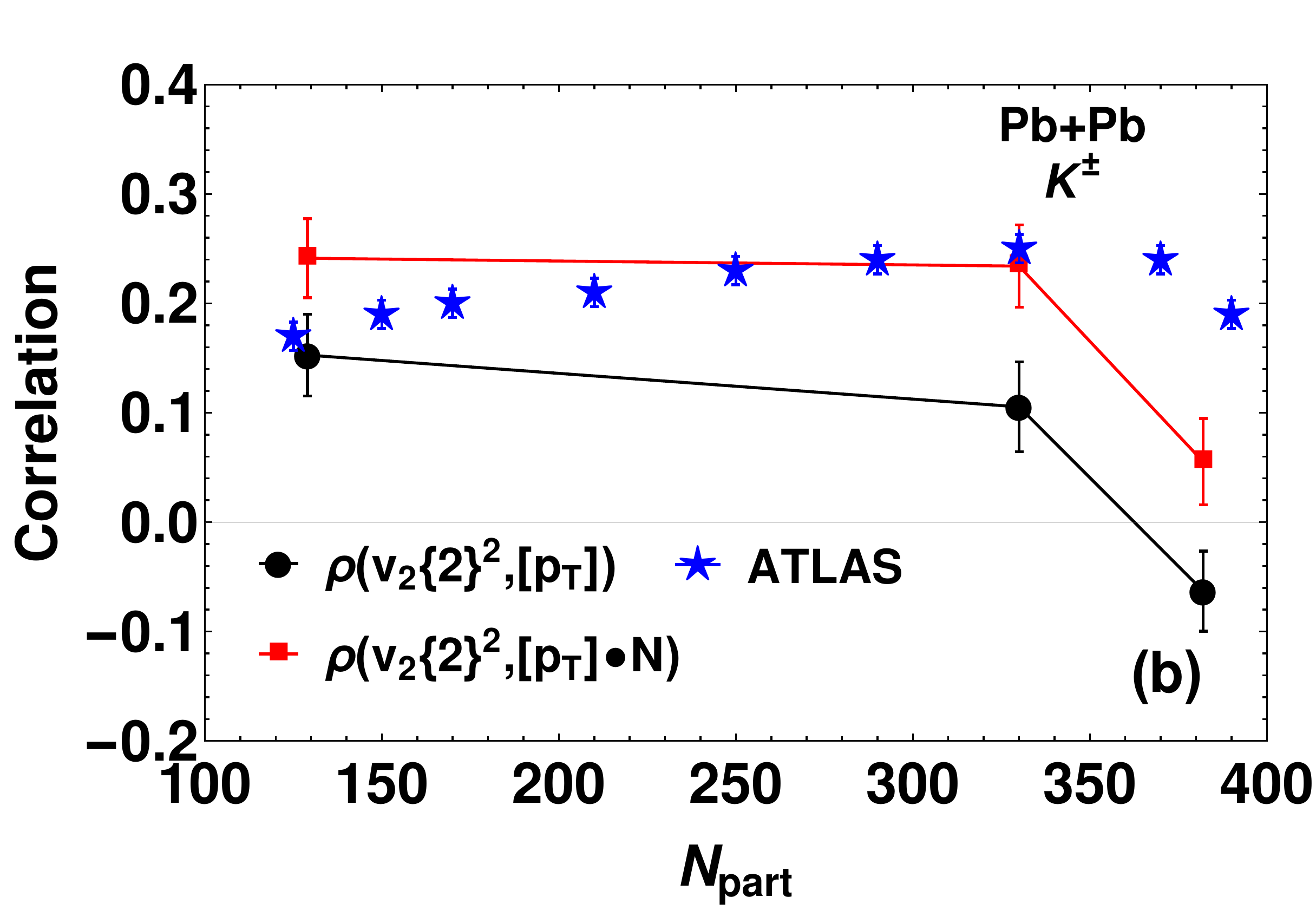} \\
			\includegraphics[scale=0.3]{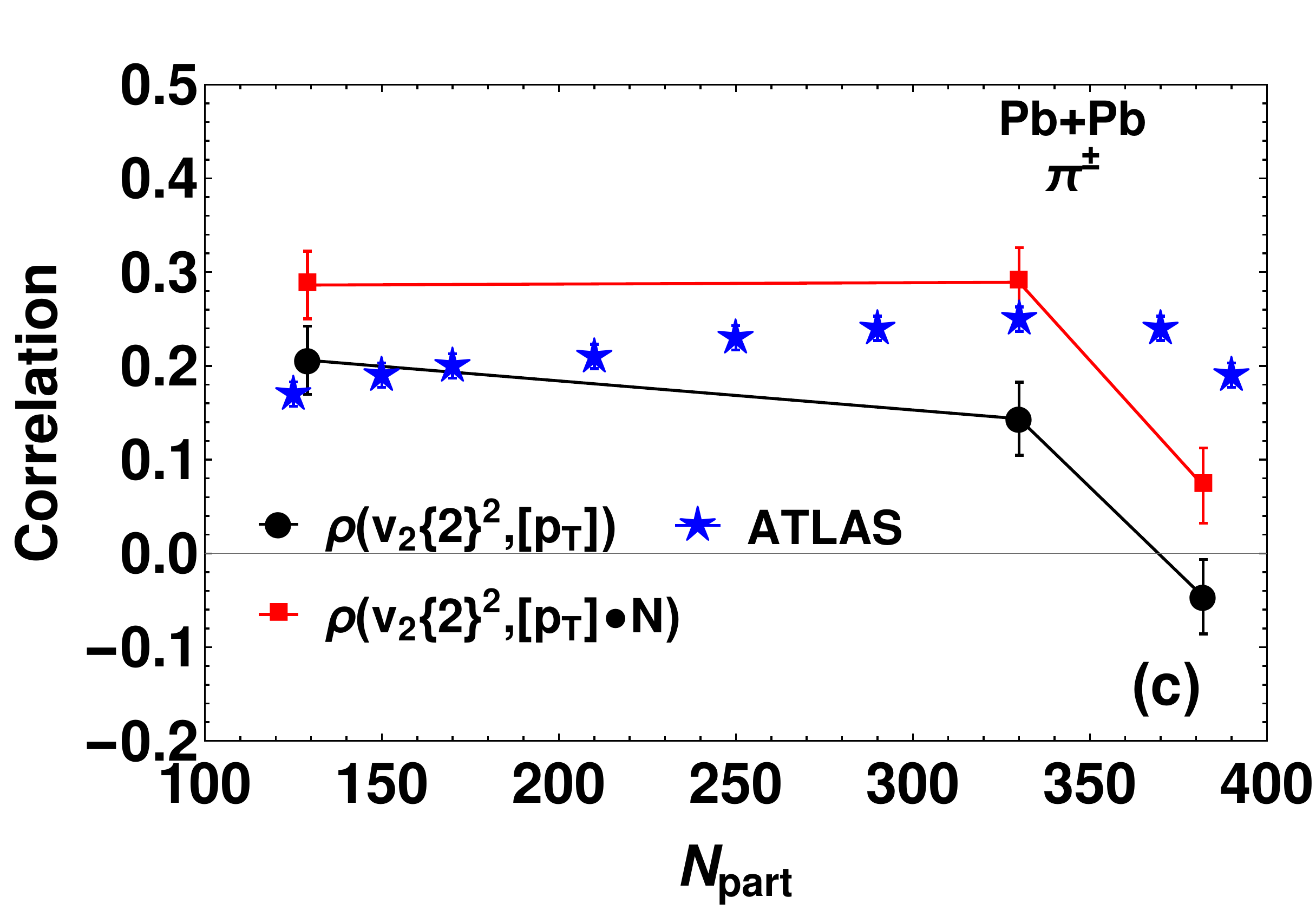} \\
  		\end{tabular}	
  		\caption{The elliptic flow-transverse momentum correlation coefficients (black dots) and the
 partial correlation coefficients (red squares) in Pb+Pb collisions for protons (panel (a)), kaons (panel (b)), and pions (panel (c)). The experimental points (blue stars) correspond to all charged particles (all panels). } 
  	\label{fig:id}
%  \end{center}
\end{figure}
\begin{figure}[ht]
%  	\begin{center}
  		\begin{tabular}{c}
  			\includegraphics[scale=0.3]{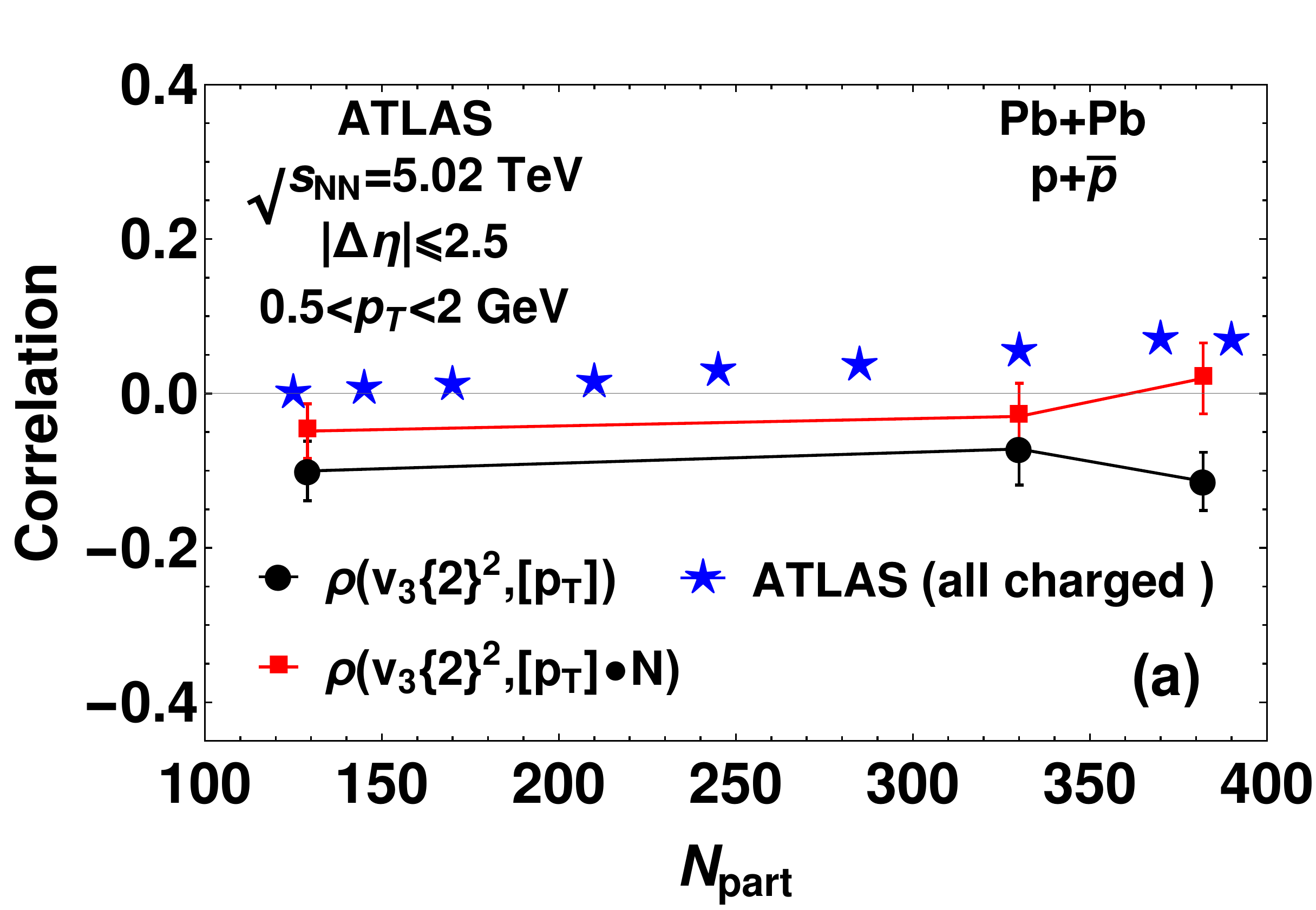} \\	       				\includegraphics[scale=0.3]{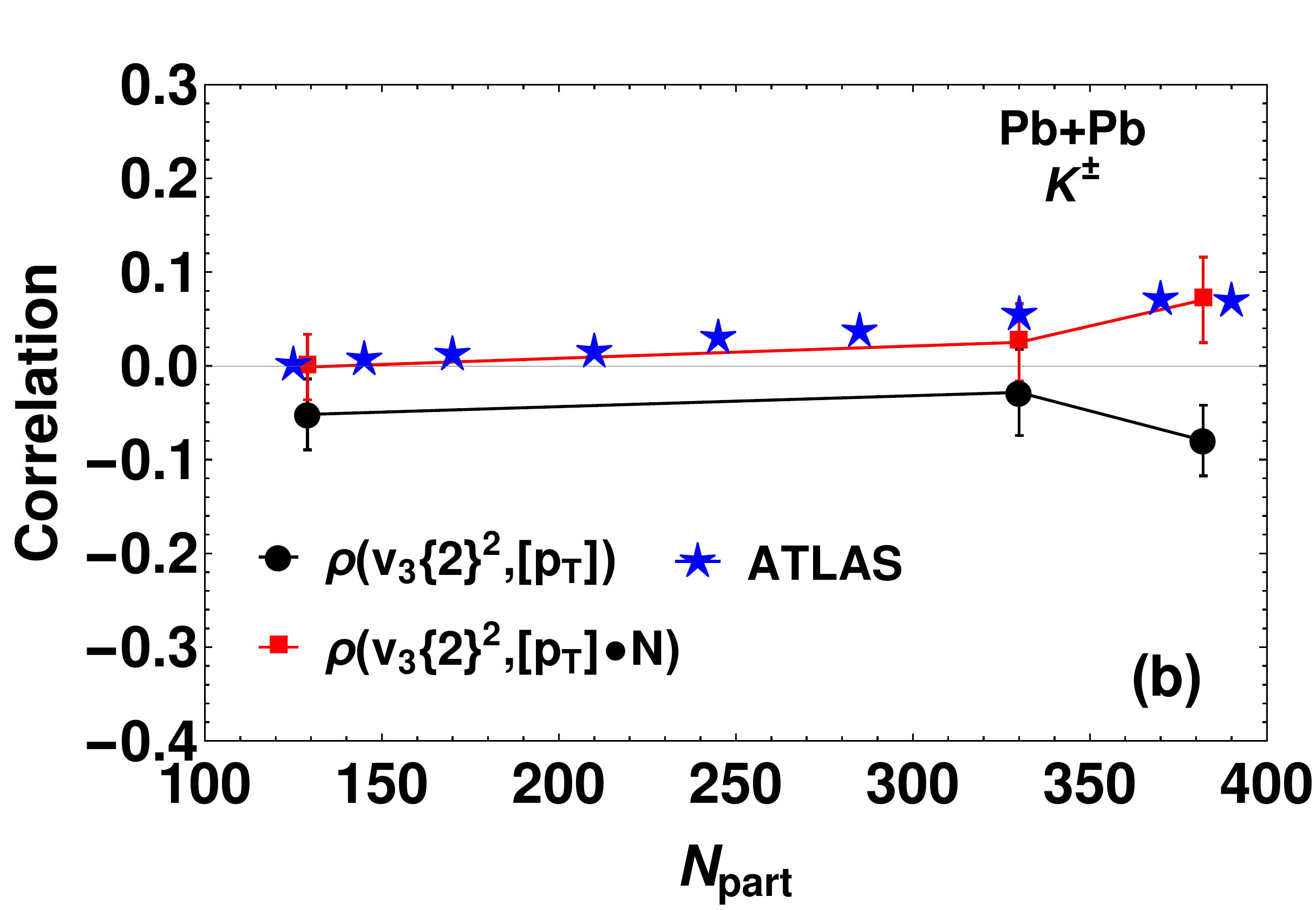} \\
			\includegraphics[scale=0.3]{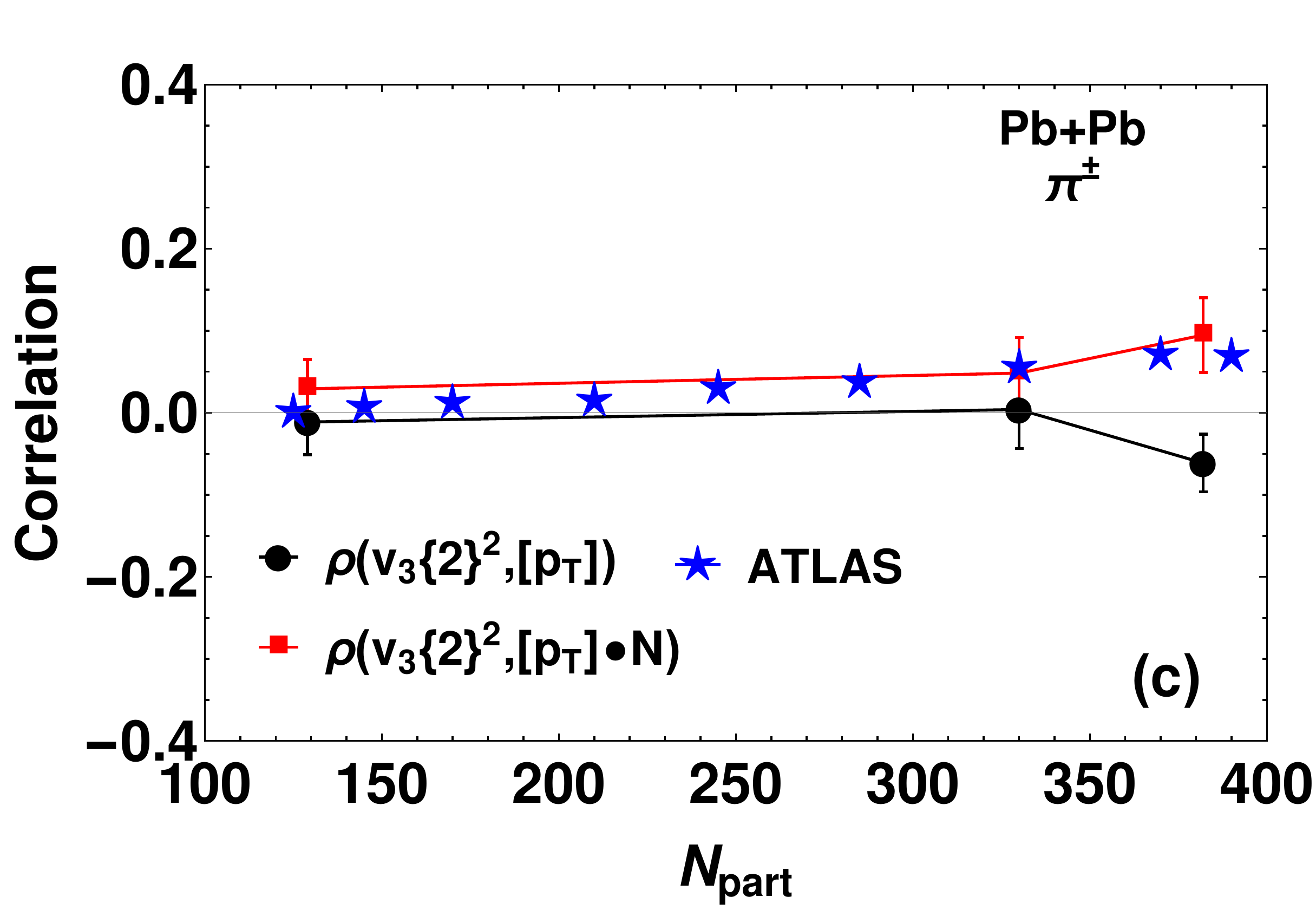} \\
  		\end{tabular}	
  		\caption{ Same as in Fig. \ref{fig:id} but for the correlation coefficient of the triangular flow with the transverse momentum. } 
  	\label{fig:idv3}
%  \end{center}
\end{figure}
The correlation between the harmonic flow and the average transverse momentum could depend on the transverse momentum cuts used for the calculation of the flow coefficients. First,
because
 the harmonic flow coefficients depend on the transverse momentum in a nonmonotonous way and second, due to an increasing contribution from mini-jets for higher $p_T$.
With increasing $p_T$ the harmonic flow-transverse momentum correlation coefficient increases (Fig. \ref{fig:ptcut}). This effect appears both in experimental data and in  simulation results. 
The flow-momentum correlation coefficient can be measured separately for different particle species. In Figs. \ref{fig:id} and \ref{fig:idv3} are presented results for the partial correlation coefficient for protons, kaons, and pions. 
The correlation coefficient becomes smaller with increasing particle mass.

%\subsection{p+Pb collisions}

\begin{figure}[t!]
%  	\begin{center}
  		\begin{tabular}{c}
  			\includegraphics[scale=0.42]{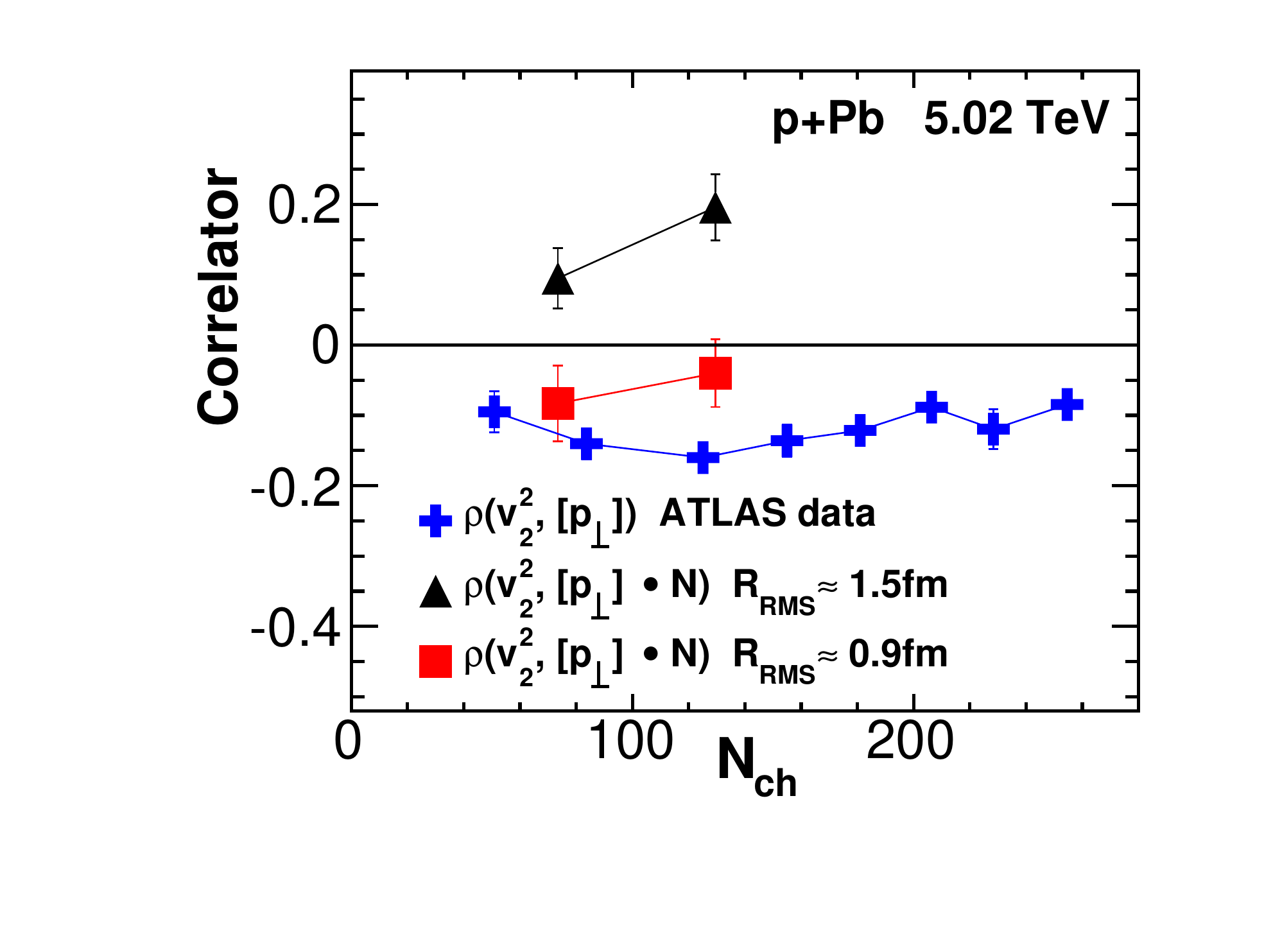} 	       						
  		\end{tabular}		
  		\caption{The elliptic flow-transverse momentum correlation coefficient in p+Pb collisions as a function of the number of  charged particles ($p_T>0.5GeV$, $|\eta|<2.5$). Two schemes for the initial state,  with two different average transverse sizes of the initial fireball, $R_{RMS}\simeq 1.5$fm (black triangles) and $R_{RMS}\simeq 0.9$fm (red squares), are compared. ATLAS Collaboration data are represented by blue crosses.} 
  	\label{fig:pPb}
%  \end{center}
\end{figure}

\begin{figure}[t!]
%  	\begin{center}
  		\begin{tabular}{c}
  			\includegraphics[scale=0.42]{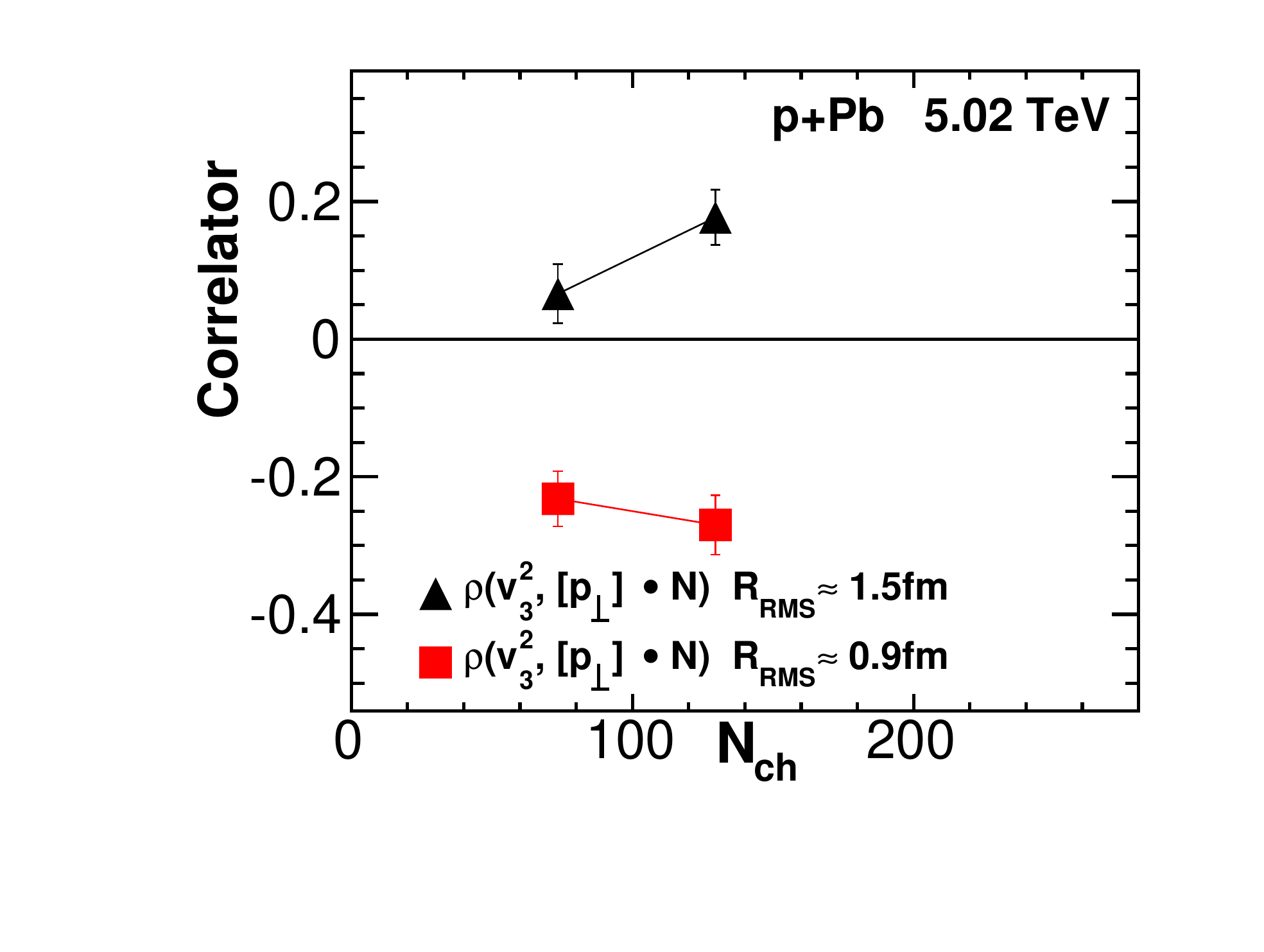} 	       						
  		\end{tabular}		
  		\caption{Same as Fig. \ref{fig:pPb} but for the triangular flow  (no experimental data available).} 
  	\label{fig:pPb3}
%  \end{center}
\end{figure}

The flow-momentum correlation can be measured also in p+Pb collisions. This 
measurement is interesting as it may give some insight on the initial state. 
Two version of the Glauber model for the initial state are used,
 the standard one, with deposition of entropy at the positions of the participant nucleons and, the second version of the model,  
with deposition of entropy in between the participant 
nucleons \cite{Bzdak:2013zma}.
The two versions of the model  give  different rms transverse sizes of the initial fireball. 
For the centralities considered in this work, the first model gives $R_{RMS}\simeq 1.5$fm and the second one $R_{RMS}\simeq 0.9$fm. The flow-momentum correlation coefficient $\rho(v_n\{2\}^2, [p_T])$ is predicted to have a different sign in the two scenarios \cite{Bozek:2016yoj}.

We present results for  the partial correlation coefficient $\rho(v_n^2\{2\},[p_T]\bullet N)$ in p+Pb collisions at $\sqrt{s_{NN}}=5.02$~TeV (Figs. \ref{fig:pPb} and \ref{fig:pPb3}). The
 change from the standard correlation coefficient  $\rho(v_n^2\{2\},[p_T])$ 
is small both for the elliptic and triangular flow. In particular, the sign of the correlation coefficient is not changed in the two scenarios for the initial state. A comparison of the calculation with experimental results on the correlation coefficient between the harmonic flow and transverse momentum favors the compact source scenario. Interestingly, also  the values  of the harmonic flow coefficients, of the average  transverse momentum, and the femtoscopy radii   are  better predicted   in the compact source scenario \cite{Bozek:2013uha,Bozek:2013df}.   The agreement with the data on  $\rho(v_n^2\{2\},[p_T])$  is worst for the
more central bin. It may indicate that  entropy fluctuations, which influence  the multiplicity and the fireball shape  in the most central p+Pb collisions, are not correctly implemented in the model. It would be interesting to confront  predictions of other models of initial state and hydrodynamic simulations on flow-transverse momentum correlations with the data. Another interesting point would be to compare the predictions of hydrodynamic and cascade models in p+Pb collisions.

\section{Estimators for flow-momentum correlations}

\begin{figure}[ht]
%  	\begin{center}
  		\begin{tabular}{c}
  			\includegraphics[scale=0.4]{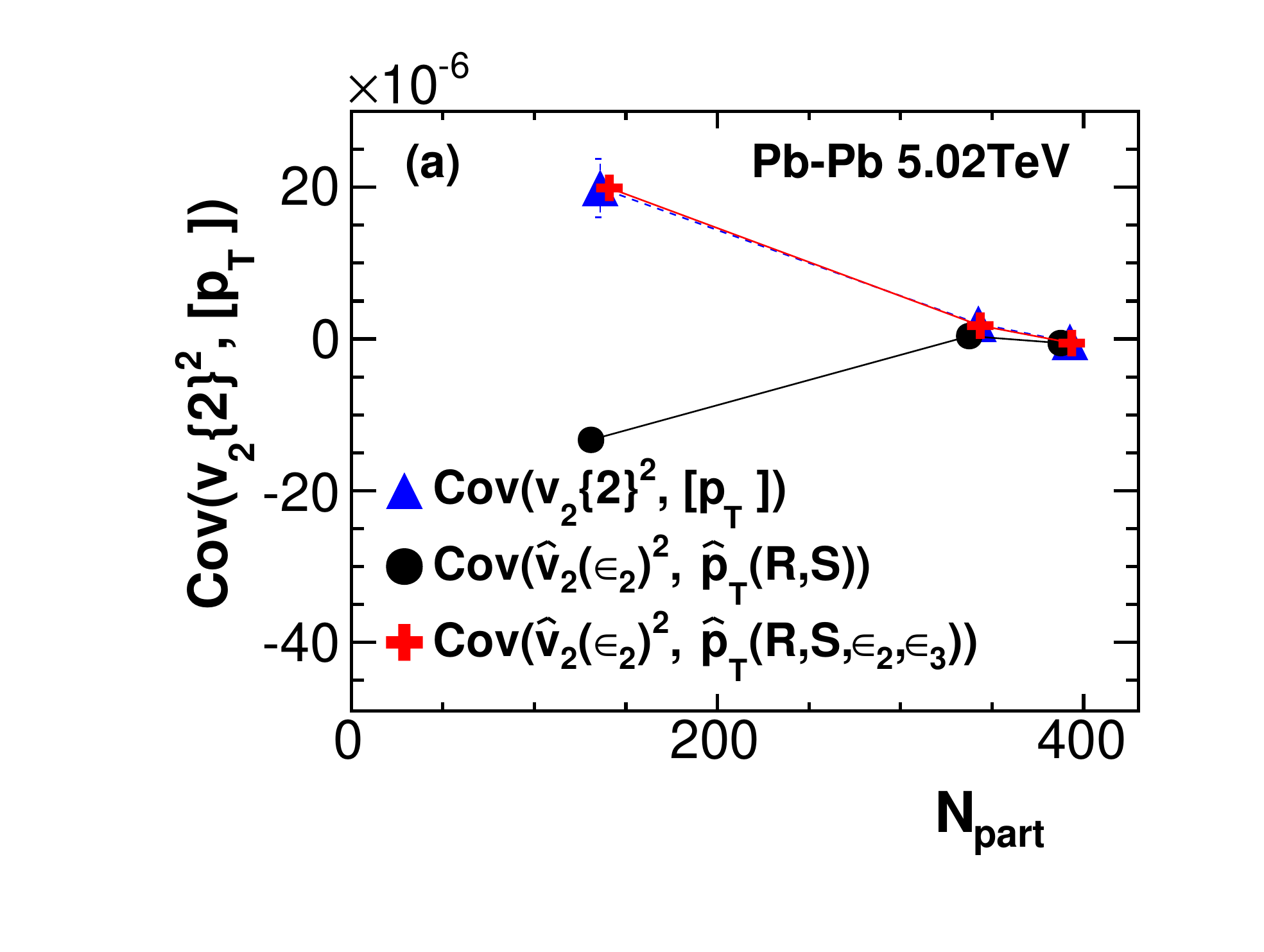} \\	       				\includegraphics[scale=0.4]{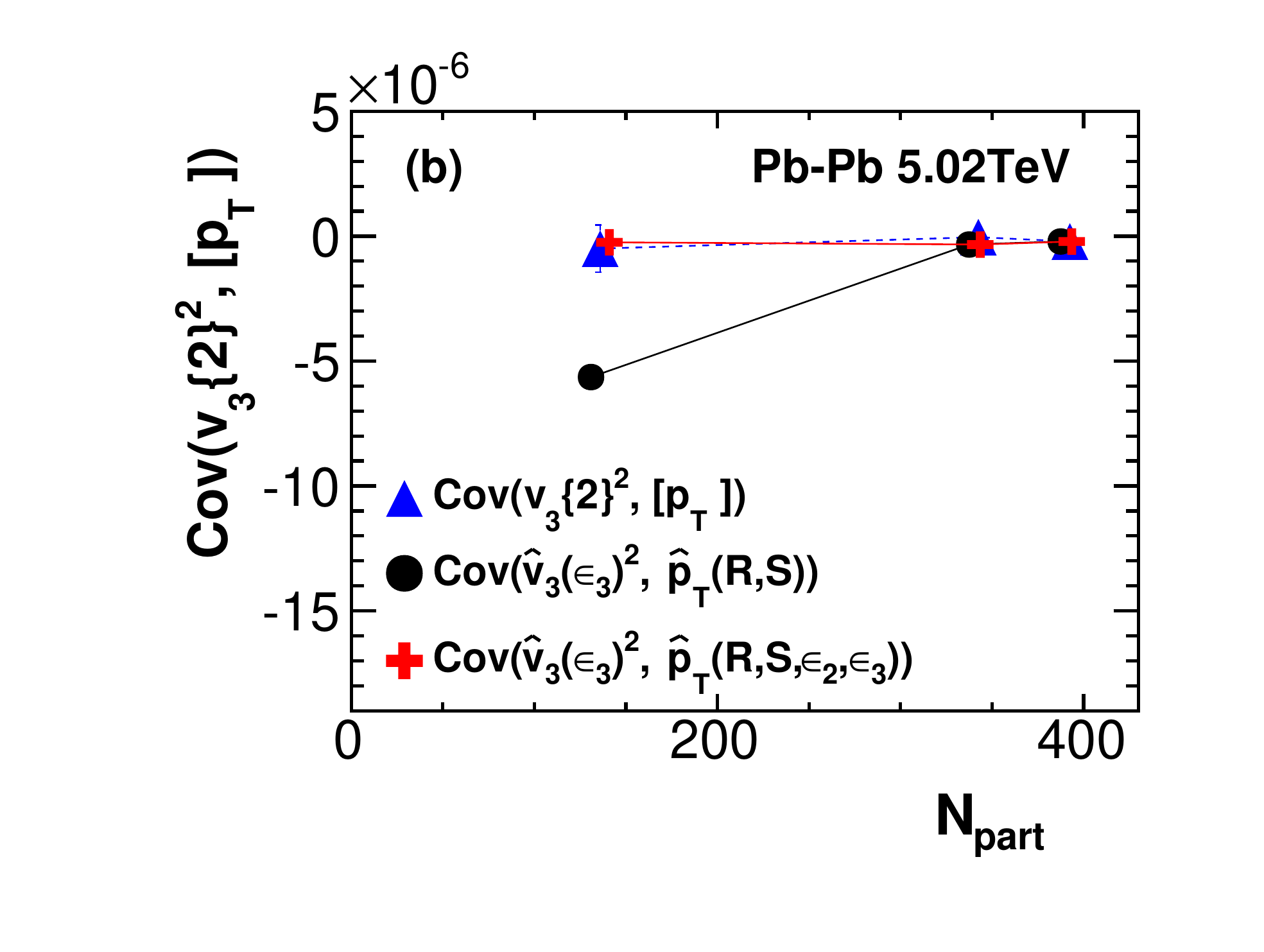} \\
  		\end{tabular}	
  		\caption{The covariance of the harmonic flow with the transverse momentum in Pb+Pb collisions as a function of the number of participant nucleons. 
The blue triangles represent the results of the hydrodynamic simulation, the black dots  represent the covariance predicted using the estimator (\ref{eq:pred1}) of the final flow harmonic and transverse momentum, the red crosses represent the covariance from the improved ansatz (\ref{eq:pred2}) for the estimator of the transverse momentum. Panels (a) and (b) present results for the elliptic and triangular flows respectively.   } 
  	\label{fig:covpt}
%  \end{center}
\end{figure}

\begin{figure}[ht]
%  	\begin{center}
  		\begin{tabular}{c}
  			\includegraphics[scale=0.4]{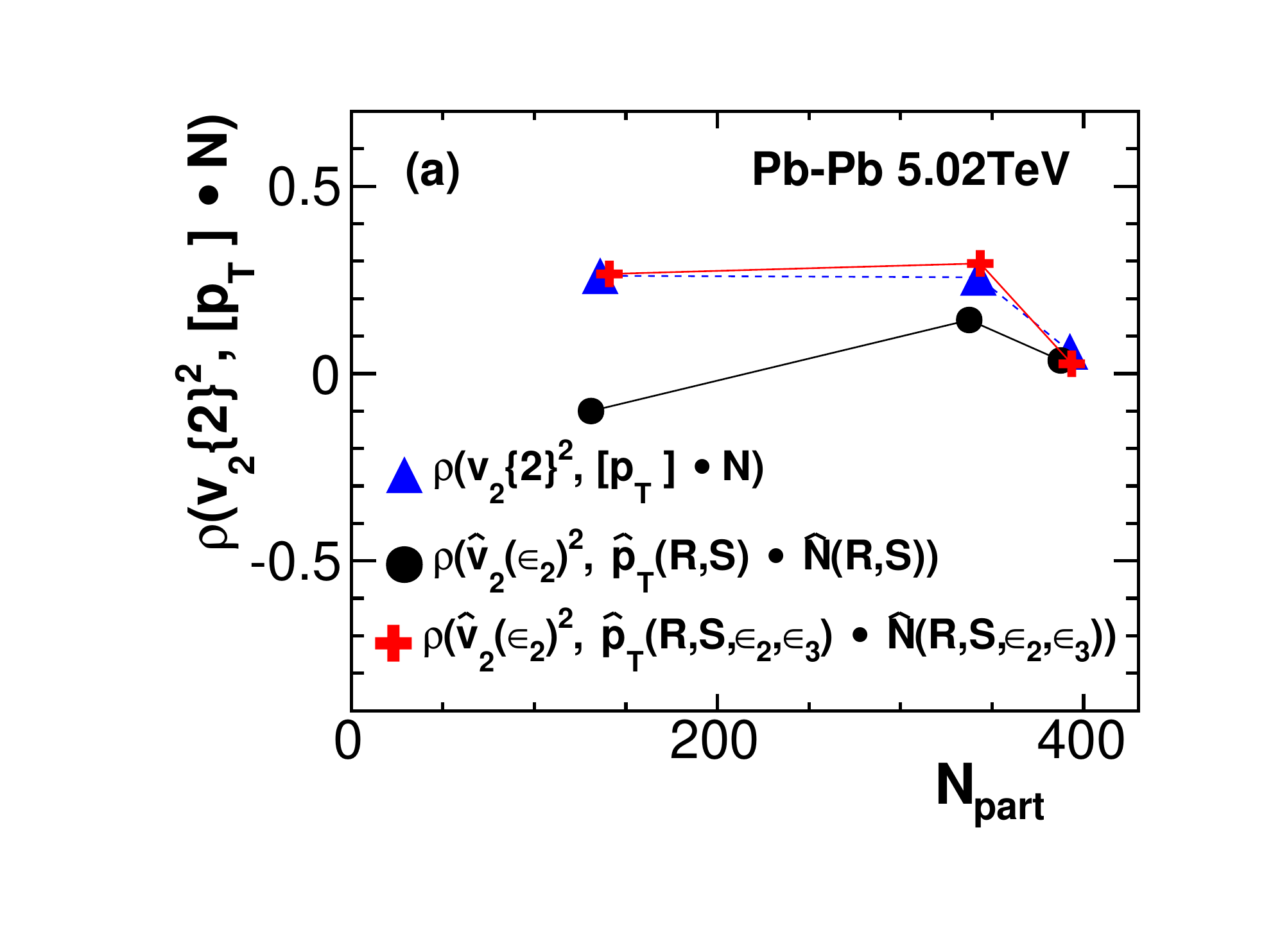} \\	       				\includegraphics[scale=0.4]{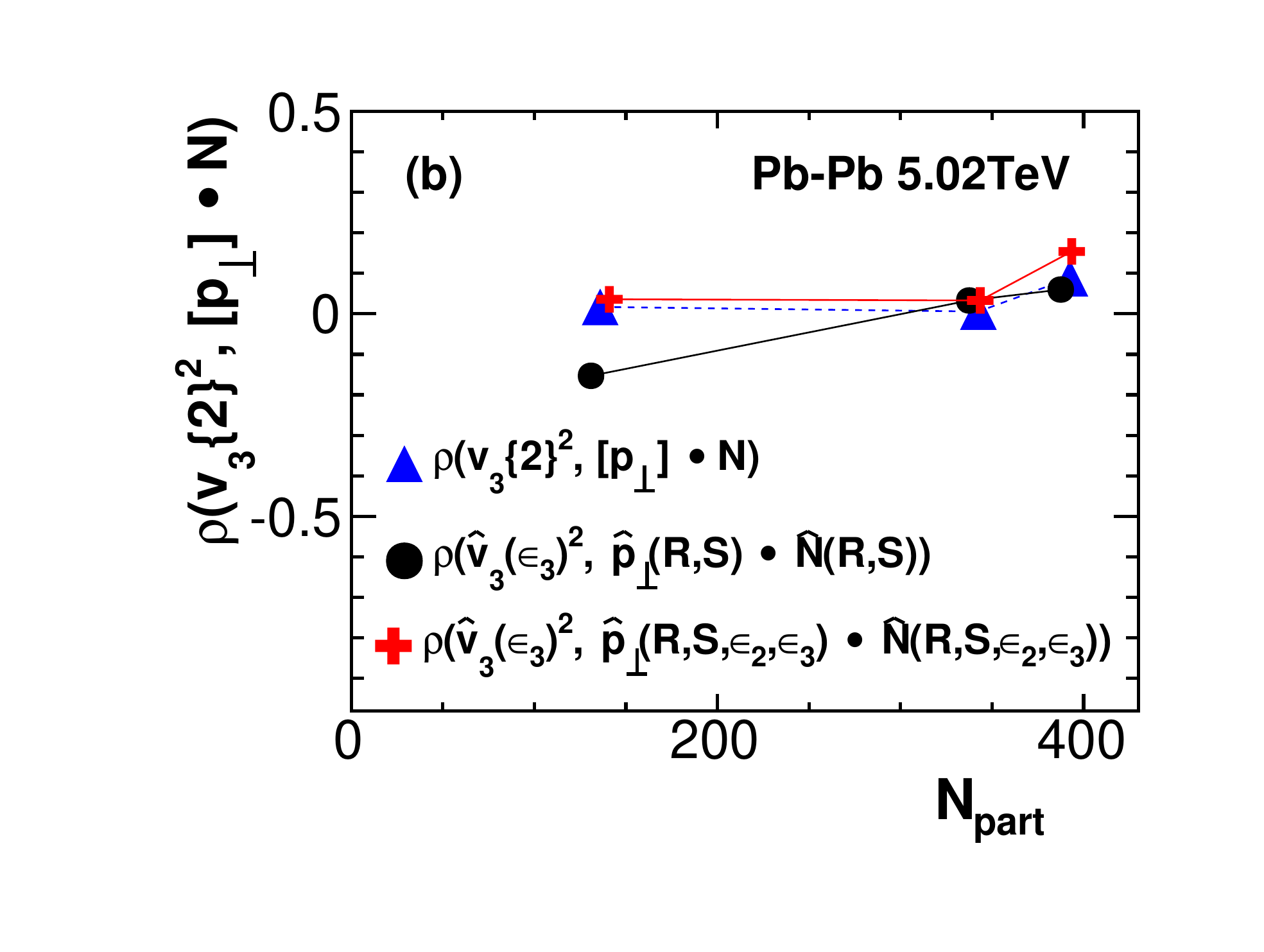} \\
  		\end{tabular}	
  		\caption{The partial correlation coefficient  of the harmonic flow with the transverse momentum in Pb+Pb collisions as a function of the number of participant nucleons. 
The blue triangles represent the results of the hydrodynamic simulation, the black dots  represent the covariance predicted using the  estimator (\ref{eq:pred1})
 of the  final flow harmonic and transverse momentum,
 the red crosses represent the covariance from the 
 improved ansatz  (\ref{eq:pred2}) 
for the estimator of the transverse momentum and multiplicity.
 Panels (a) and (b) present results for the elliptic and triangular
 flows respectively.   } 
  	\label{fig:corrvpt}
%  \end{center}
\end{figure}

\begin{figure}[ht]
%  	\begin{center}
  		\begin{tabular}{c}
  			\includegraphics[scale=0.4]{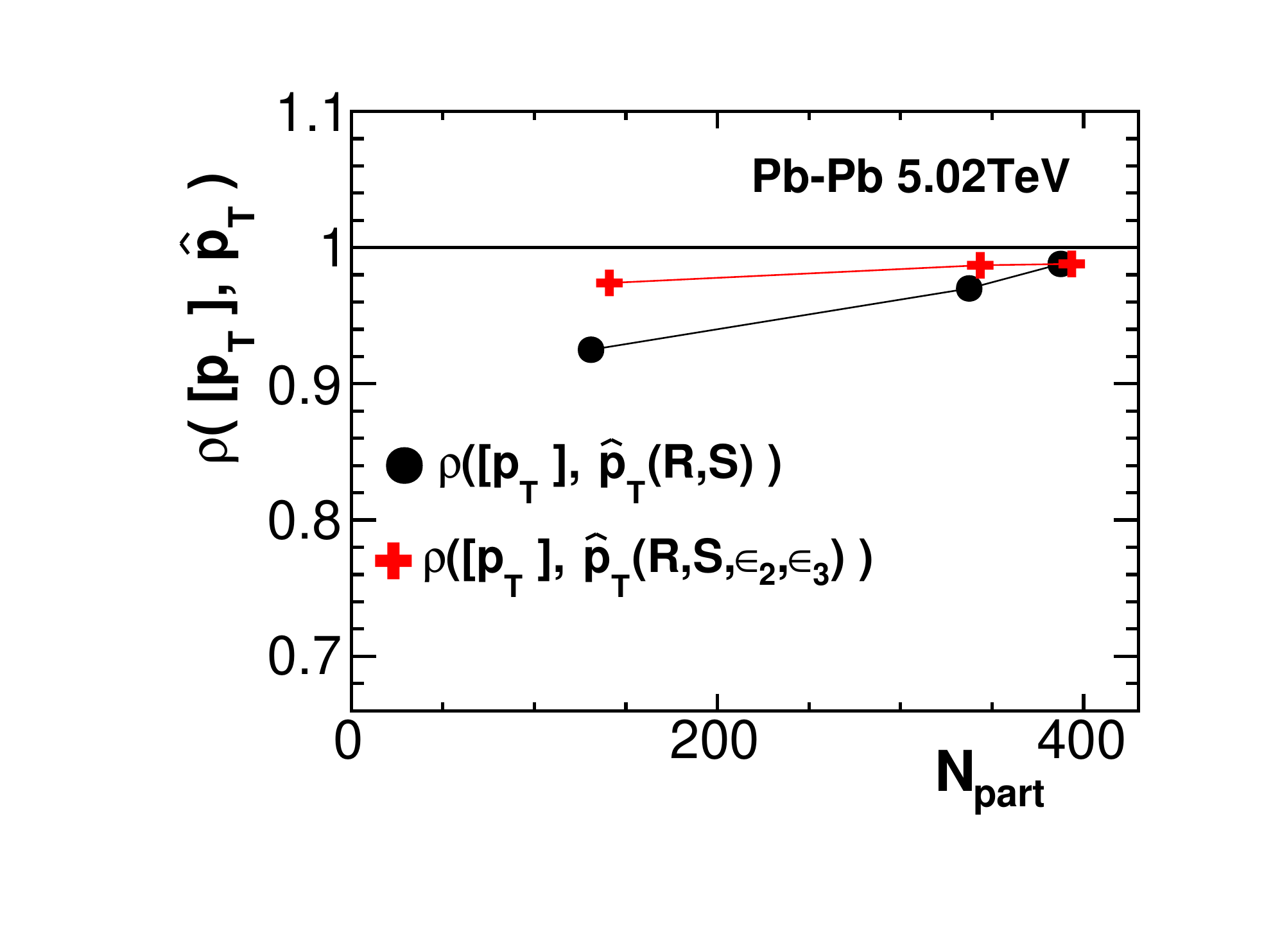} 
  		\end{tabular}	
  		\caption{The correlation coefficient  of the average transverse momentum $[p_T]$ and its predictor $\hat{p}$ in Pb+Pb collisions as a function of the number of participant nucleons. 
 The black dots  represent results for the  predictor  depending on the 
initial transverse size and entropy (\ref{eq:pred1}),  the red crosses correspond to the predictor with initial eccentricities added to the estimator formula (\ref{eq:pred2}).   } 
  	\label{fig:pprep}
%  \end{center}
\end{figure}

The correlation between the harmonic flow and the transverse momentum of final charged 
hadrons results from a hydrodynamic response applied to a given ensemble of initial conditions in event by event evolution. In this paper we consider a linear response to initial conditions. The linear response estimators  from the initial eccentricities is  a good approximation  of the final harmonic  flow
\cite{Gardim:2011xv,Qiu:2011iv,Niemi:2012aj}. 
The average transverse momentum in an event is largely determined by the initial transverse size of the fireball  \cite{Broniowski:2009fm}. Additional corrections to the predictor for  transverse momentum 
come from the initial entropy and eccentricities \cite{Mazeliauskas:2015efa,Bozek:2017elk}.

In the following we study to predictors for the final global observables $[p_T]$, $v_2\{2\}^2$, $v_3\{2\}^2$, and $N$. The first ansatz for the predictors is
\begin{eqnarray}
\hat{p}(R_{RMS}, S) & = &\langle [p_t] \rangle + a_p (R_{RMS}-\langle R_{RMS}\rangle ) 
+b_p (S-\langle S \rangle ) \nonumber \\
\hat{N}(R_{RMS}, N) & =  & a_N (R_{RMS}-\langle R_{RMS}\rangle ) 
+b_N S \nonumber \\
\hat{v}_{2}\{2\}^2 (\epsilon_2) & = & k_2 \epsilon_2^2  \nonumber \\
\hat{v}_{3}\{2\}^2 (\epsilon_3) & = & k_3 \epsilon_3^2   \ \ ,
\label{eq:pred1}
\end{eqnarray}
where the initial transverse rms radius in an event  is 
\begin{equation}
R_{RMS} = \left [ \frac{\int rdr d\phi \  r^{2} s(r,\phi )}{\int rdr d\phi   s(r,\phi )} \right]^{1/2}
\end{equation}
and the initial entropy is 
\begin{equation}
S= \int rdr d\phi \   s(r,\phi)  \  .
\end{equation}
Note that the linear predictor for the average transverse momentum is constructed as a linear relation for the deviation from the average $[p_T]-\langle [p_T]\rangle$. The  average $\langle [p_T]\rangle$ itself depends on scales imposed on the dynamics, freeze-out temperature and hydrodynamic evolution time, not only on the initial conditions. 
The coefficients  ($a_p$, \dots , $k_3$) of the  linear relation \ref{eq:pred1}
are adjusted to minimize the sum of square  deviations between the prediction
 and the actual value of the global observable for  events corresponding
 to a given centrality class.

The covariance between the harmonic flow and transverse momentum $Cov(v_n\{2\}^2,[p_T])$
in shown in Fig. \ref{fig:covpt}. The covariance obtained using the hydrodynamic simulations (blue triangles) is compared  to the covariance of flow and transverse momentum obtained using the estimator \ref{eq:pred1} (black dots).
For central collisions the covariance obtained using the predictors reproduces
 the 
hydrodynamic results. For semiperipheral collisions
 the deviation is significant.
The same is true for the partial correlation coefficient
 (Fig. \ref{fig:corrvpt}), which involves also the predictor for the final
 multiplicity. 

  In the improved ansatz,  eccentricities $\epsilon_n^2$ are
added to the estimator formula
 \begin{eqnarray}
\hat{p}(R_{RMS}, S) & = &\langle [p_t] \rangle + a_p (R_{RMS}-\langle R_{RMS}\rangle ) 
+b_p (S-\langle S \rangle ) \nonumber \\
& & + c_p (\epsilon_2^2-\langle \epsilon_2^2 \rangle) + d_p (\epsilon_3^2
-\langle \epsilon_3^2 \rangle) \nonumber \\
\hat{N}(R_{RMS}, N) & = & a_N (R_{RMS}-\langle R_{RMS}\rangle ) 
+b_N S \nonumber \\
&& + c_N (\epsilon_2^2-\langle \epsilon_2^2 \rangle) + d_N (\epsilon_3^2
-\langle \epsilon_3^2 \rangle) \nonumber \\
\hat{v}_{2}\{2\}^2 (\epsilon_2) & = & k_2 \epsilon_2^2  \nonumber \\
\hat{v}_{3}\{2\}^2 (\epsilon_3) & = & k_3 \epsilon_3^2   \ \ .
\label{eq:pred2}
\end{eqnarray}
The flow-transverse momentum  covariance $Cov(v_n\{2\}^2,[p_T])$ 
from the hydrodynamic simulations is well described using the improved ansatz (red crosses in Fig. \ref{fig:covpt}).  Also the partial correlation coefficient is fairly well described using the improved ansatz (Fig. \ref{fig:corrvpt}).
The essential part of the improvement comes from the inclusion of the eccentricities $\epsilon_n^2$ in the ansatz for the average transverse momentum.
Fig. \ref{fig:pprep} presents the correlation coefficient between the transverse momentum $[p_T]$ and its predictor $\hat{p}$. The inclusion of the eccentricities in the predictor (\ref{eq:pred2}) for $\hat{p}$ increases the correlation $\rho([p_T],\hat{p_T})$. It should be noted that estimators of the initial size in non-central collisions, other than the rms transverse radius, have been discussed as determining the transverse expansion
\cite{Bhalerao:2005mm,Lacey:2014wqa}.

\section{Effect of control variable for other observables}

\begin{figure}[t!]
%  	\begin{center}
  		\begin{tabular}{c}
  			\includegraphics[scale=0.29]{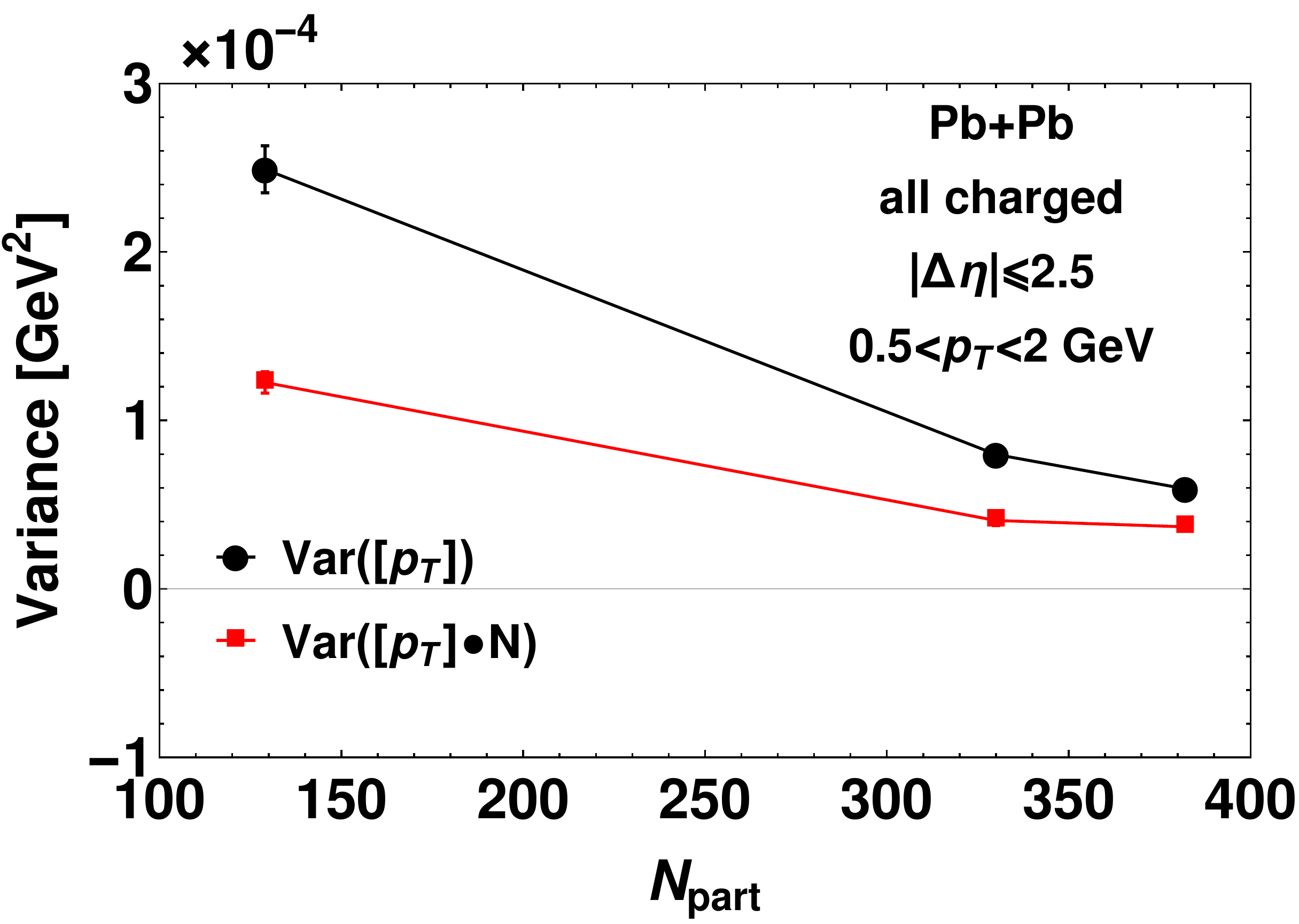} 	       						
  		\end{tabular}		
  		\caption{ The  variance and partial variance of the average transverse momentum in Pb+Pb collisions as a function of the number of participant nucleons.} 
  	\label{fig:varn}
%  \end{center}
\end{figure}

Another observable discussed in heavy ion collisions is the variance of the average transverse momentum \cite{Broniowski:2009fm,Adams:2003uw,Abelev:2014ckr,Gavin:2003cb}. In the hydrodynamic model transverse momentum fluctuations reflect the fluctuations of the initial volume \cite{Broniowski:2009fm} and the violence of the collective transverse expansion \cite{Ollitrault:1991xx}.
For broad centrality bins, multiplicity fluctuations are important. Multiplicity fluctuations influence significantly the variance of the 
average transverse momentum. The partial variance of the transverse momentum with respect to the multiplicity is significantly smaller than the standard variance. This observation should be kept in mind when comparing simulations and data in centrality bins corresponding to different widths of multiplicity distributions.

\begin{figure}[ht]
%  	\begin{center}
  		\begin{tabular}{c}
  			\includegraphics[scale=0.29]{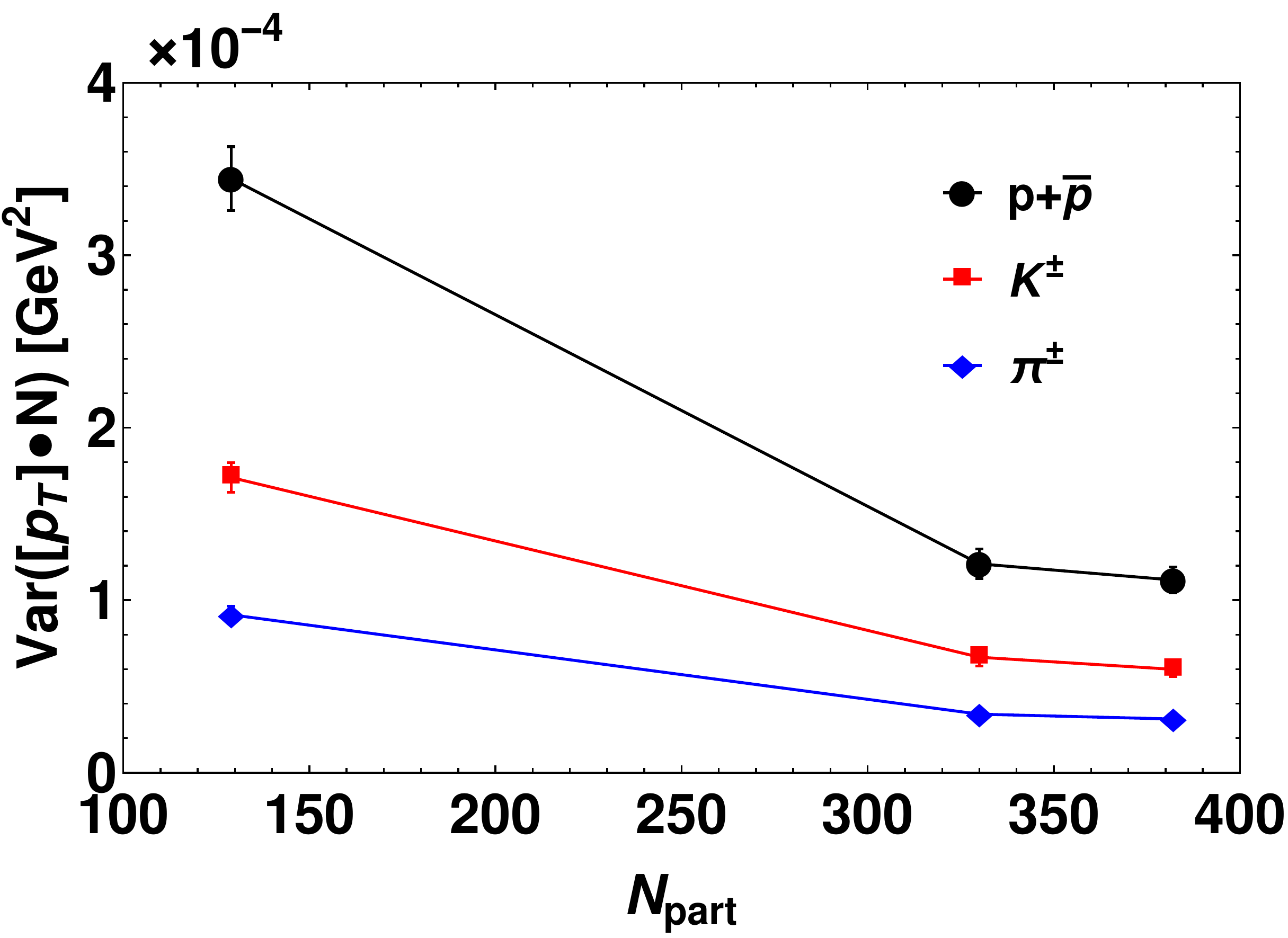} \\	       	  		\end{tabular}	
  		\caption{ The dynamical variance of the event by event average transverse momentum $Var([p_T])$ for protons (black circles), kaons (blue diamonds), and pions (red squares) in Pb+Pb collisions as a function of the number of participants nucleons.} 
  	\label{fig:varid}
%  \end{center}
\end{figure}

\begin{figure}[ht]
%  	\begin{center}
  		\begin{tabular}{c}
  			\includegraphics[scale=0.29]{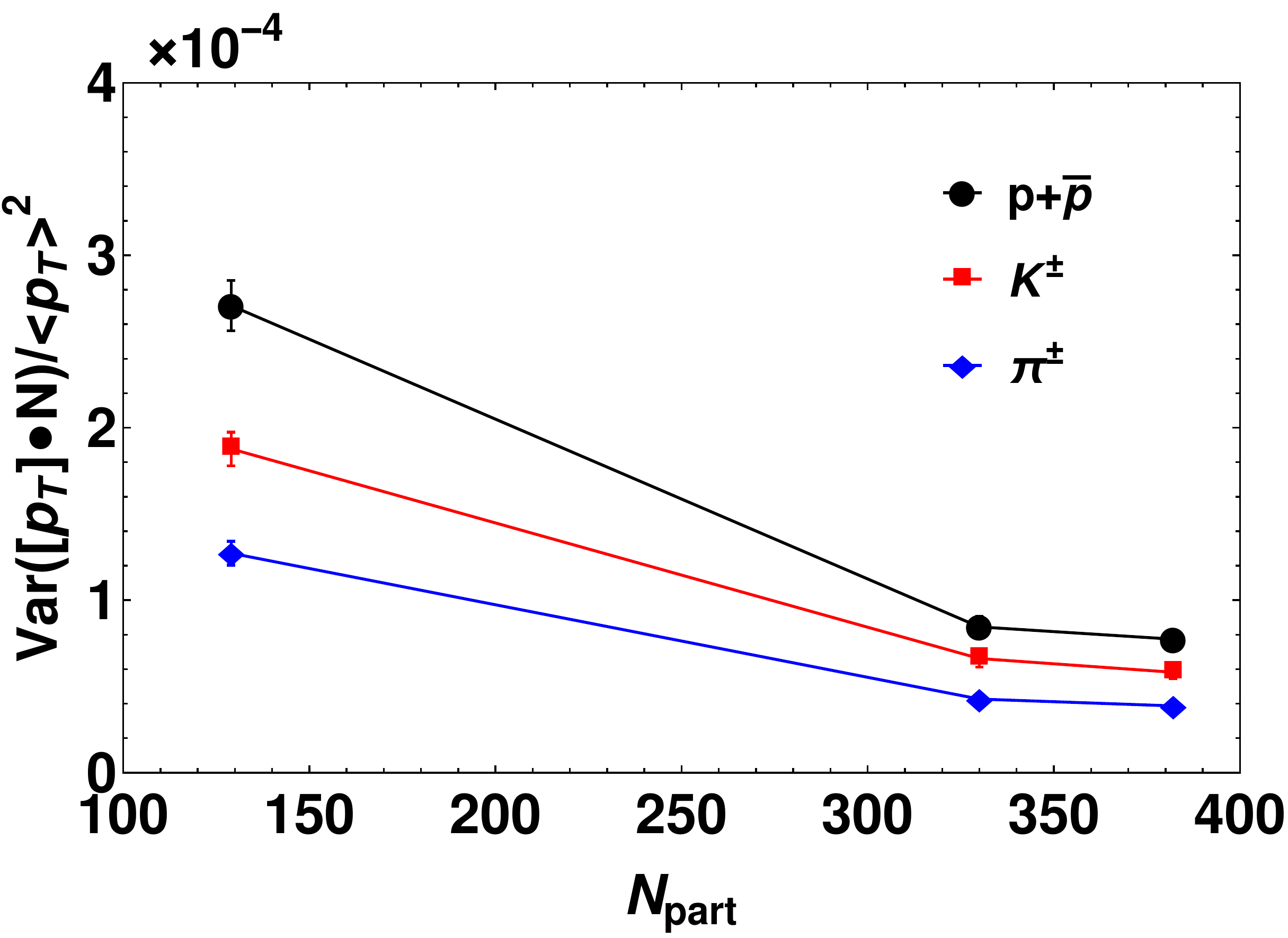} \\	       	  		\end{tabular}	
  		\caption{ The dynamical variance of the event by event  average transverse momentum scaled by the square of the corresponding  average of the transverse momentum $Var([p_T])/\langle  p_T \rangle^2$ for protons (black circles), kaons (blue diamonds), and pions (red squares) in Pb+Pb collisions as a function of the number of participants nucleons.} 
  	\label{fig:varidsc}
%  \end{center}
\end{figure}

In Fig. \ref{fig:varid} is shown the dynamical  variance of the average transverse momentum for identified particles. Again, to correct for the multiplicity fluctuations, we present results for the particle variance of the average transverse momentum. The  event by event fluctuations of the average transverse momentum are larger for massive particles. Part of this dependence may be due to the increase of the average transverse momentum of emitted particles with particle mass. 
 The  variance scaled by the square of average  transverse momentum 
$\frac{Var([p_T])}{\langle p_T \rangle^2}$ shows still a clear dependence on particle mass (Fig. \ref{fig:varidsc}). With increasing particle mass the contribution of collective flow increases with respect to the thermal momentum.

\begin{figure}[ht]
%  	\begin{center}
  		\begin{tabular}{c}
  			\includegraphics[scale=0.29]{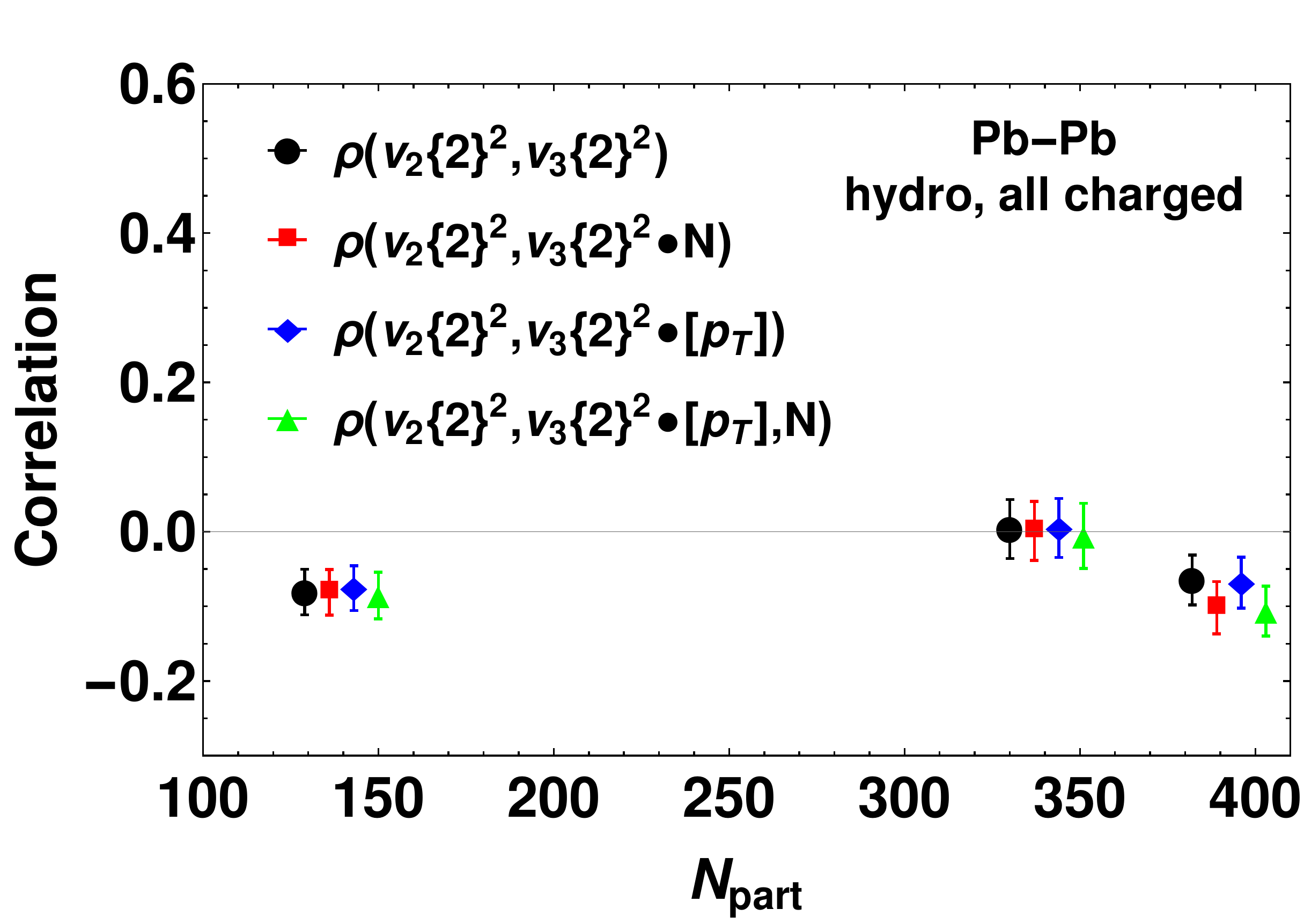} \\	       	  		\end{tabular}	
  		\caption{ The correlation coefficient between the elliptic and the triangular flow $\rho(v_2\{2\}^2,v_3\{2\}^2)$ (black dots), and the partial correlation coefficients $\rho(v_2\{2\}^2,v_3\{2\}^2\bullet N)$ (red squares),
$\rho(v_2\{2\}^2,v_3\{2\}^2\bullet [p_T])$ (blue diamonds), and
$\rho(v_2\{2\}^2,v_3\{2\}^2\bullet N, [p_T])$ (green triangles) as  functions of
the number of participant nucleons.} 
  	\label{fig:v2v3}
%  \end{center}
\end{figure}

As a further example, we  study the partial correlation for cumulants
 \cite{Bilandzic:2010jr} of harmonic flows with corrections for control 
variables $[p_T]$ and $N$. We present results for the correlation 
coefficient between the elliptic and triangular flows 
$\rho(v_2\{2\}^2,v_3\{2\}^2)$. In Fig. \ref{fig:v2v3} we compare the 
standard  correlation coefficient and the partial correlation coefficients with respect to multiplicity $\rho(v_2\{2\}^2,v_3\{2\}^2\bullet N)$, to transverse momentum $\rho(v_2\{2\}^2,v_3\{2\}^2\bullet [p_T])$, and to both control variables  
$\rho(v_2\{2\}^2,v_3\{2\}^2\bullet N,[p_T])$. 
The corrections due to correlations  of flow cumulants 
with control variables are negligible.

\section{Conclusions}

Correlations between the harmonic flow  coefficients and the average transverse momentum are studied for relativistic collisions at $\sqrt{s_{NN}}=5.02$~TeV. Hydrodynamic
 model results are compared to experimental data of the ATLAS Collaboration \cite{Aad:2019fgl}. Hydrodynamic simulations reproduce fairly well the measurements for central and semi-central Pb+Pb collisions.  In p+Pb collisions the hydrodynamic model with initial condition corresponding to a compact, small-sized source reproduces  qualitatively the measurement, while the standard Glauber model initial conditions lead a wrong sign of the correlation coefficient. 

A novelty in the analysis is the incorporation of corrections 
due to correlations to a control variable, the multiplicity.
Hydrodynamic simulations are performed in centrality bins with relatively 
broad multiplicity distributions. The effect of multiplicity fluctuations
on the  correlation coefficients can be corrected using the partial correlation coefficient \cite{Olszewski:2017vyg}. The correction is sizable 
 for the correlation of the elliptic flow and transverse momentum and for the variance of the transverse momentum.

The covariance between the final harmonic flow and transverse momentum results from the hydrodynamic response on the covariance matrix of the 
initial eccentricities, rms transverse size and multiplicity. A good ansatz for the linear hydrodynamic response requires the combination of the transverse size, entropy, and eccentricities in the estimator for the final transverse momentum.

\section*{Acknowledgments}
HM would like to thank the AGH UST for the great hospitality during the course of this work on his collaboration leave, as well as his research adviser Hessamaddin Arfaei for all his support. PB thanks Derek Teaney for helpful suggestions. This research is supported by the AGH University of Science and Technology, by the Institute for Research in Fundamental Sciences (IPM), and  by the  Polish National Science Centre grant 2018/29/B/ST2/00244. 
\bibliography{../hydr}

\end{document}